%% file: main.tex
\documentclass[a4paper,11pt]{article}
\usepackage{aaskaiid}
\usepackage{hyperref}
\usepackage[nameinlink,capitalise]{cleveref}
\usepackage{amsfonts,amsmath,bm,mathtools}
\usepackage{ulem,comment,siunitx}
\DeclareSIUnit\Jy{Jy}
\DeclareSIUnit\parsec{pc}
\usepackage{tikz}
\usetikzlibrary{arrows.meta}
\usepackage{orcidlink}

\newcommand{\lcdm}{\(\Lambda\)CDM}
\newcommand{\hi}{\textsc{Hi}}

\newcommand{\dd}{\delta_{\rm D}}

\newcommand{\ii}{\mathrm{i}}

\DeclareMathOperator{\snr}{SNR}

\newcommand{\Sam}[1]{\textcolor{green}{#1}}

\newcommand{\toma}[1]{\textcolor{olive}{#1}}

\let\Gamma\varGamma
\let\Delta\varDelta
\let\Theta\varTheta
\let\Xi\varXi
\let\Pi\varPi
\let\Upsilon\varUpsilon
\let\Phi\varPhi
\let\Psi\varPsi
\let\Omega\varOmega

\title{Beyond \lcdm\ with the SKA Observatory -- I\\
Probing Gravity on Cosmological Scales}
\ShortTitle{Beyond $\varLambda$CDM with the SKAO -- I: Probing Gravity on Cosmological Scales}

\author[1,2,3,4]{S.\ Camera\orcidlink{0000-0003-3399-3574}}
\ShortName{Camera et al.}
\emailAdd{stefano.camera@unito.it}
\author[5]{C.\ Addis\orcidlink{0000-0003-4906-9634}}
\author[4]{S.L.\ Guedezounme\orcidlink{0009-0000-9200-8584}}
\author[1,2]{F.\ Montano\orcidlink{0009-0000-2520-9861}}
\author[6]{T.\ Montandon\orcidlink{0000-0002-5536-6953}}
\author[1,2]{M.\ Novara}
\author[1,2]{S.J.\ Rossiter\orcidlink{0009-0005-7953-2396}}
\author[7]{J.\ Asorey\orcidlink{0000-0002-6211-499X}}
\author[8]{S.\ Castello\orcidlink{0000-0002-1178-5371}}
\author[5,4]{C.\ Clarkson}
\author[9,10,4]{J.\ Fonseca\orcidlink{0000-0003-0549-1614}}
\author[11,12]{N.\ Frusciante\orcidlink{0000-0002-7375-1230}}
\author[13,14]{C.\ Gomes\orcidlink{0000-0001-6022-459X}}
\author[15,8]{N.\ Grimm\orcidlink{0000-0001-9602-0599}}
\author[16]{A.\ Nasirudin}
\author[9,10]{L.G.T.\ Oliveira}
\author[1,2,3]{F.\ Pace\orcidlink{0000-0001-8039-0480}}
\author[17,18,19]{Z.\ Sakr\orcidlink{0000-0002-5149-4042}}
\author[8]{M.\ Berti\orcidlink{0000-0002-7667-4883}}
\author[8]{C.\ Bonvin}
\author[16,4]{P.\ Bull}
\author[8]{E.\ Di Dio}
\author[20]{C.\ Cress}
\author[21]{C.\ Heneka}
\author[22]{H.-R.\ Kl\"ockner\orcidlink{0000-0002-0648-2704}}
\author[4]{R.\ Maartens}
\author[9,23]{C.J.A.P.\ Martins\orcidlink{0000-0002-4886-9261}}
\author[24]{D.\ Parkinson\orcidlink{0000-0002-7464-2351}}
\author[24]{C.S.\ Saraf\orcidlink{0000-0002-5149-4042}}

\affiliation[1]{Dipartimento di Fisica, Universit\`a degli Studi di Torino, 10125 Torino, Italy}
\affiliation[2]{INFN -- Istituto Nazionale di Fisica Nucleare, Sezione di Torino, 10125 Torino, Italy}
\affiliation[3]{INAF -- Istituto Nazionale di Astrofisica, Osservatorio Astrofisico di Torino, 10025 Pino Torinese, Italy}
\affiliation[4]{Department of Physics \& Astronomy, University of the Western Cape, Cape Town 7535, South Africa}
\affiliation[5]{Astronomy Unit, Department of Physics \& Astronomy, Queen Mary University of London, E1 4NS, U.K.}
\affiliation[6]{Laboratoire Univers et Particules,
CNRS \& Universit\'e de Montpellier, 34095 Montpellier, France}
\affiliation[7]{Departamento de F\'isica Te\'orica, Centro de Astropart\'iculas y F\'isica de Altas Energ\'ias (CAPA), Universidad de Zaragoza, 50009 Zaragoza, Spain}
\affiliation[8]{D\'epartement de Physique Th\'eorique and Center for Astroparticle Physics, Universit\'e de Gen\`eve, CH-1211 Geneva 4, Switzerland}
\affiliation[9]{Instituto de Astrof\'isica e Ci\^encias do Espa\c{c}o, Universidade do Porto CAUP, 4150-762 Porto, Portugal}
\affiliation[10]{Faculdade de Ci\^{e}ncias, Universidade do Porto, 4169-007 Porto, Portugal}
\affiliation[11]{Dipartimento di Fisica ``E.\ Pancini'', Universit\`a degli Studi di Napoli ``Federico II'', 80126 Napoli, Italy}
\affiliation[12]{INFN -- Istituto Nazionale di Fisica Nucleare, Sezione di Napoli, 80126 Napoli, Italy}
\affiliation[13]{Centro de F\'isica das Universidades do Minho e do Porto, Faculdade de Ciências, Universidade do Porto, 4169-007 Porto, Portugal}
\affiliation[14]{OKEANOS, Universidade dos A\c cores, 9900-140 Horta, Portugal}
\affiliation[15]{Institute of Cosmology \& Gravitation, University of Portsmouth, Portsmouth PO1 3FX, U.K.}
\affiliation[16]{Jodrell Bank Centre for Astrophysics, The University of Manchester, Manchester M13 9PL, U.K.}
\affiliation[17]{Instituto de F\'isica Te\'orica UAM-CSIC, Universidad Aut\'onoma de Madrid, 28049 Madrid, Spain}
\affiliation[18]{Institut de Recherche en Astrophysique et Plan\'etologie (IRAP), Universit\'e de Toulouse, CNRS, UPS, CNES, 31400 Toulouse, France}
\affiliation[19]{Faculty of Sciences, Universit\'e Saint-Joseph de Beyrouth, Beirut BP-11514, Lebanon}
\affiliation[20]{Department of Mathematical Sciences, University of South Africa, Roodepoort 1709, South Africa}
\affiliation[21]{Institut f\"ur Theoretische Physik, Universit\"at Heidelberg, 69120 Heidelberg, Germany}
\affiliation[22]{Max-Planck-Institut f\"ur Radioastronomie, 53121 Bonn, Germany}
\affiliation[23]{Centro de Astrof\'{\i}sica da Universidade do Porto, 4150-762 Porto, Portugal}
\affiliation[24]{Korea Astronomy and Space Science Institute, Daejeon 34055, Republic of Korea}

\let\Gamma\varGamma
\let\Delta\varDelta
\let\Theta\varTheta
\let\Xi\varXi
\let\Pi\varPi
\let\Sigma\varSigma
\let\Upsilon\varUpsilon
\let\Phi\varPhi
\let\Psi\varPsi
\let\Omega\varOmega

\abstract{
General relativity (GR) is currently the best description of the gravitational interaction at our disposal and is one of the foundations of the concordance cosmological model. For as much as we know that GR is not the final theory of gravitation---we still lack an understanding of its fundamental, quantum nature---it has demonstrated a remarkable success in describing observed phenomena and predicting effects that have later been confirmed by laboratory experiments or astronomical observations. Since gravity is extremely weak compared to the other three fundamental interactions, it has so far been tested with exquisite precision only in the strong-field regime. On the immense scales of the cosmos, on the other hand, the gravitational field is extremely weak and spacetime curvature is almost negligible. But crucially, it is on these scales that we see hints at the need for exotic components, such as dark matter and  dark energy. The question of whether they really exist or their presence is but an artifact of the incompleteness of our understanding of gravity on cosmological scales then naturally arises. It is therefore paramount to test the validity of GR on these scales, either to further confirm its robustness or to detect deviations that could lead us to the formulation of a more general and conclusive theory of gravitation. To this purpose, the SKA Observatory is especially suited, thanks both to the enormous volumes it will probe, and to the variety and complementarity of cosmological observables that its surveys will make available to us.
}

\begin{document}
\maketitle
\tableofcontents

\section{Introduction}
Over the past decades, the so-called concordance cosmological model, \lcdm, has provided us with a remarkably successful description of the Universe on large scales. It rests on two pillars: general relativity (GR), to describe the gravitational interaction, and the Standard Model of Particle Physics, accounting for the content of the Universe as well as for the interplay between it and the other three known fundamental interactions. In addition to such pillars, it is then necessary to hypothesise the presence of a yet-unknown cold dark matter (CDM) component, and a cosmological constant ($\Lambda$) as the driver of the observed late-time accelerated expansion of the cosmos. Among the remarkable successes of the \lcdm\ model are the reproduction of the observed expansion history and the cosmic microwave background (CMB) anisotropies, and of the statistical properties of the large-scale distribution of galaxies, with impressive precision \citep[see e.g.][]{Planck:2018vyg,eBOSS:2020yzd}. However, \lcdm\ faces a number of conceptual and observational challenges that motivate the exploration of alternatives. The focus of this chapter is thus particularly on testing the first pillar, i.e.\ the validity of GR as the correct theory of gravity on cosmological scales.

From a theoretical standpoint, the cosmological constant problem remains one of the deepest puzzles in modern physics. The observed value of $\Lambda$, corresponding to an energy density of roughly $10^{-120}$ in Planck units, is inexplicably small compared to the natural scales predicted by quantum field theory \citep{Weinberg:1988cp,Padmanabhan:2002ji}. This extreme fine-tuning, together with the so-called coincidence problem---why are the density of matter and that associated to \(\Lambda\) comparable today?---has led to a long-standing search for a dynamical explanation of cosmic acceleration. Cosmologists have tried to model the accelerated expansion with two alternative approaches. The first assumes that GR is valid on all (macroscopic) scales and then introduces a significant fraction of dark energy, intended as an evolving component whose present-day value is almost indistinguishable from a simple cosmological constant \citep[see][and references therein, for a comprehensive review]{ParticleDataGroup:2024cfk}. Alternatively, the acceleration can be modelled as an emergent effect from modifications of gravity on cosmological scales \citep[see e.g.][]{2012PhR...513....1C}.

To get some intuition, let us consider \cref{fig:four_fields}. Cosmological observables essentially probe the four fields presented there and their interrelations. On cosmological scales, we see (left panel) how GR predicts a precise relation between the content of the Universe and the distortions its presence causes on space---namely, the relativistic Poisson equation---as well as how space and time distortions are related. The continuity and Euler equations instead arise from the conservation of the stress-energy tensor, with cosmic acceleration being due to the cosmological constant or a dynamical dark energy component. Evidence of any modification to this standard scenario (right panel) will guide us to its source, and the advent of next-generation surveys offers unprecedented opportunities to test these scenarios \citep[for a review on the topic, see also][]{2019LRR....22....1I}.
\input{plots/Fig1}

In this context, the SKA Observatory (SKAO) will play a transformative role in probing the nature of dark energy and gravity. With its large sky coverage, wide redshift range, and exquisite sensitivity, the SKAO will provide us with a diverse and complementary set of cosmological observables, which could also be analysed jointly with more traditional cosmological observables at microwave and optical/near-infrared wavelengths.
By measuring the expansion rate 
and the angular diameter distance
through baryon acoustic oscillations
and redshift-space distortions (RSDs), and the growth rate of structures, 
which depends directly on the underlying theory of gravity, the SKAO can tightly constrain deviations from GR.
Moreover, the various cosmological probes available to us thanks to the SKAO will allow for powerful, model-independent consistency tests \citep{Camera:2015yqa,Baker:2019gxo}.

The synergistic potential of the SKAO with traditional optical/near-infrared surveys for cosmology
will further enhance its ability to trace both geometry and the growth of cosmic structures, contributing to lifting degeneracies and  uncovering possible scale-dependent deviations from GR. The coming decades will hence be pivotal in determining whether cosmic acceleration arises from an evolving dark energy component or from a breakdown of GR on cosmological scales. Accordingly, the present chapter is structured to reflect these two possibilities. After having introduced the main observables in \cref{sec:obs}, \cref{sec:rel_effects} will outline how cosmological surveys with the SKAO will be able to test the validity of GR on cosmological scales. Conversely, \cref{sec:mg} will focus on the capabilities of the SKAO in detecting signatures of modified gravity. \Cref{sec:additional} will present additional probes to test gravity with the SKAO. Future perspectives, also in the view of possible upgrades of the instrument and technological advancements, will be discussed in \cref{sec:SKA2}. Finally, conclusions will be drawn in \cref{sec:conclusions}.

\section{Target observables and summary statistics}\label{sec:obs}

In the study of the cosmic large-scale structure (LSS), two cosmological probes have been identified by the scientific community as particularly promising: galaxy clustering and cosmic shear. The former scrutinises the statistical properties of fluctuations in the observed number counts of galaxies, whereas the latter extracts information from the distorted shapes of distant galaxy images, in the regime of weak gravitational lensing deflections. At the state of the art, they are, respectively, the primary targets of spectroscopic surveys performed e.g.\ by the Dark Energy Spectroscopic Instrument \citep[DESI,][]{2022AJ....164..207D,2025arXiv250314745D} or imaging surveys like the Legacy Survey of Space and Time \citep[LSST,][]{2019ApJ...873..111I,2018arXiv180901669T} at the Vera C.\ Rubin Observatory. Each observable is a powerful source of information in itself, but it is well-known that their combination opens up transformational possibilities, such as detecting deviations from GR on cosmological scales \citep{2004PhRvD..70d3009H,2004ApJ...600...17B}. It is precisely for this reason that the European Space Agency’s flagship, the \textit{Euclid} satellite, was designed in the first place \citep{2025A&A...697A...1E}. 

In this context, the field of radio cosmology will be revolutionised by the SKAO \citep{Maartens:2015mra,SKA:2018ckk}, thanks to the new observables that it will make available and their cross-correlations with clustering and shear measured by traditional means.
Among such new observables, neutral hydrogen (\hi) intensity mapping stands out. It is a revolutionary new way of mapping the LSS by measuring the aggregate line emission from all of the gas and galaxies within a relatively large volume element, thus obtaining high-fidelity representations of the underlying matter distribution \citep{Bharadwaj2001,Loeb2008}. 
Currently\Sam{,} surveys carried out at so-called precursor and pathfinder facilities \citep{2011PASA...28..215N,2016mks..confE..32S,2017A&A...598A.104S} are already demonstrating the enormous potential of future SKAO observational campaigns for cosmology.

Finally, even gravitational wave (GW) observations can be exploited to probe the LSS, e.g.\ with LIGO, Virgo, and Kagra \citep{LVK_4OR2025}, the Einstein Telescope \citep[ET,][]{ET_2025}, or the Laser Interferometer Space Antenna \citep[LISA,][]{LISA_2023}. We thus expect plenty of varied data sets to be available in the upcoming years to investigate the cosmic LSS.

We shall now briefly outline the two main cosmological fields under consideration: the density contrast of a tracer \(t\) of the underlying LSS, \(\Delta_t\), and the weak gravitational lensing effect of cosmic shear, \(\gamma\). 
For more details on clustering and lensing in an SKAO context, we refer the reader to the following other chapters of the present science book: \citet[][for \hi-line galaxies]{Nasirudin01.2026.SKA}, \citet[][for radio-continuum galaxies]{Asorey01.2026.SKA}, \citet[][for \hi\ intensity mapping]{Wolz01.2026.SKA}, \citet[][for radio weak lensing]{Harrison01.2026.SKA}, and \citet[][for cross-correlations among the aforementioned probes]{Harrison02.2026.SKA}. 

\subsubsection*{Galaxy clustering and \hi\ intensity mapping}\label{sec:clustering}
Fluctuations in galaxy number density trace perturbations in the underlying matter density via multiple effects. In a fully general-relativistic treatment, various sub-dominant contributions arise as a consequence of the propagation of photons in an inhomogeneous universe and we consider these below.
In the linear regime, where cosmological perturbations are small, the number density contrast can be schematically written as the sum of various contributions \citep[see][]{Yoo:2009au, Bonvin:2011bg, Challinor:2011bk, 2012PhRvD..85b3504J},
\begin{equation}\label{eq:Delta}
\Delta_t=\Delta_t^{\rm(den)}+\Delta_t^{\rm(vel)}+\Delta_t^{\rm(len)}+\Delta_t^{\rm(rel)}\;,
\end{equation}
where \(t\) labels any given galaxy population tracing the underlying LSS. The first term captures the effect of matter density perturbations, given by the galaxy bias \(b_t\) times the matter density contrast \(\delta\). The second one includes contributions due to peculiar velocities, which perturb the observed redshift. The dominant of these velocity contributions is the well-known RSD effect \citep{Kaiser:1987qv,Zaroubi_1993}, $\Delta^{\rm(RSD)}\subset \Delta_t^{\rm(vel)}$. However, $\Delta_t^{\rm(vel)}$ also contains a relativistic Doppler term that is typically not considered in RSD analyses. The third term describes the increase/depletion of galaxy number counts caused by lensing magnification. Increase is due to more sources falling within the detection threshold thanks to their flux being amplified by lensing, whilst depletion may happen because of the stretch of the distorted solid angle, which may make sources at the edge of an observed patch fall out of the area. Finally, the last term collects all other contributions, collectively referred to as `relativistic'. Among such terms are an integrated Sachs-Wolfe effect, time delay, and gravitational redshift.

Typically, only the dominant terms \(\Delta_t^{\rm(den)}=b_t\,\delta\) and \(\Delta^{\rm(RSD)}\) are included in standard analyses from spectroscopic surveys \citep[e.g.][]{2025JCAP...09..008A},\footnote{Note that \(\Delta^{\rm(RSD)}\) is sample-independent and, as such, has no \(t\) subscript.} with \(\Delta_t^{\rm(len)}\) appearing in the most recent analyses of photometric surveys \citep{2023A&A...675A.189D,2025arXiv250313632D,2025arXiv250505821S}.
Finally, note that \hi\ intensity mapping can also be described within the same formalism, where now \(t=\hi\) labels the fluctuation in the mean temperature brightness of the 21-cm sky, \(T_\hi=(1+\Delta_\hi)\,\bar T_\hi\), and it is worth noting that in this case the lensing term identically vanishes, because surface brightness is conserved \citep[see][for an exhaustive treatment of relativistic effects in \hi\ intensity mapping]{2013PhRvD..87f4026H}.

\subsubsection*{Cosmic shear and cross-correlations}\label{sec:shear}
The other main cosmological observable of the LSS is cosmic shear. Shear is one of the two components into which the amplification matrix of gravitational lensing can be decomposed, in the weak lensing regime. Convergence is the other component.
Shear acts on a galaxy image by stretching the image along one direction and contracting it in the orthogonal direction. Technically, it is characterized by a \(2\times2\) traceless and symmetric tensor, whose two independent components, \(\gamma_{i=1,2}\), can then be conveniently combined into a complex, spin-2 shear field, \(\gamma=\gamma_1+\ii\,\gamma_2\). Analogously to CMB polarisation, we can introduce shear parity-eigenstates, called \(E\)- and \(B\)-modes, with the main advantages being: dealing with spin-0---and thus rotationally invariant---quantities; and allowing for a clean separation between cosmological signal and systematic effects---as, at linear order, \(B\)-modes cannot be generated in a parity-symmetric universe.

Cosmic shear is especially relevant in testing gravity because it is sourced by the Weyl potential, \(\Upsilon\coloneqq(\Psi+\Phi)/2\), where \(\Psi\) is the gravitational potential that, together with the other metric potential \(\Phi\), forms the pair of gauge-invariant potentials (see \cref{fig:four_fields}).
Lensing hence allows us to test modified gravity theories where the two potentials are not the same, unlike in GR (in the absence of anisotropic stress). Furthermore, cosmic shear is highly complementary to galaxy clustering, as the latter is mainly sensitive to \(\Psi\) alone, in the standard scenario.
For this reason, weak lensing analyses usually consider the shear signal measured from galaxies jointly with their distribution, thus performing a so-called 3\texttimes2-point (`pt', for short) analysis, which includes shear-shear, galaxy-galaxy, and shear-galaxy correlations. It is worth remarking that relativistic effects contribute to cosmic shear only at second order \citep{2010PhRvD..81h3002B}, whilst they impact lensing convergence already at linear order \citep{2014MNRAS.443.1900B}, thus potentially affecting shear-galaxy correlations \citep{2018JCAP...06..008G}. 

More generally, we shall refer to analyses of one single observable (e.g.\ shear-shear) as `auto-correlations' and to those where the focus is on the interplay between two or more observables (like shear-galaxy) as `cross-correlations'.

\subsection{\(N\)-pt correlation functions and poly-spectra}\label{sec:correlators}
Cosmological perturbations are stochastic and, as such, we can study their statistical properties. In particular, it is customary to define summary statistics in terms of moments of their probability distributions. Depending on whether we study perturbations in position (also, configuration) space or in some dual (e.g.\ Fourier) space, we shall refer to them as \(N\)-pt correlation functions or poly-spectra, respectively.

Cosmological perturbations are defined as fluctuations on top of a homogeneous and isotropic background. Hence, they average to zero, but their variance encodes information about their primordial distribution, as well as about the processes they underwent during the formation of the LSS \citep[e.g.][]{Peebles:1980yev}. The (co)variance of a cosmological field \(X\) at comoving position \(\bm r\) and a (possibly different) field \(Y\) at position \(\bm r'\) is called the 
2-pt correlation function, which we denote by \(\xi_{XY}(\bm r,\bm r')\coloneqq\langle X(\bm r)\,Y(\bm r')\rangle\). 
In some cases, it is preferable to work in Fourier space, because  on linear scales different Fourier modes evolve independently of each other. Thus, we define the 
power spectrum as the correlator of two perturbations in Fourier space, \(\langle X(\bm k)\,Y(\bm k')\rangle\eqqcolon(2\,\pi)^3\,\dd(\bm k+\bm k')\,P_{XY}(\bm k)\), where \(\dd\) is the Dirac-delta distribution and \(\bm k\) and \(\bm k'\) are Fourier wave-vectors---having used, with an abuse of notation, the same symbol for a configuration-space function and its Fourier transform. 

A key assumption here is statistical homogeneity (translational invariance) in the local region where we are correlating our \(N\) points \citep{Zaroubi_1993,Scoccimarro_2015}. However, as we correlate points at wide separations, this symmetry is broken. First, it is broken angularly, as we leave the plane-parallel limit, and as such, due to RSDs the points that we are correlating have different underlying statistics. Secondly, it is broken radially, due to the evolution of cosmological fields on the past light-cone. In 
Fourier poly-spectra we can introduce perturbative corrections in terms of the ratio between the distance to the two points, \(\bm d\), and their separation, \(\bm s=\bm r-\bm r'\), to account for some of this lost information when we assume translational invariance. These are the so-called wide-angle effects \citep{Reimberg_2016,Castorina_2018,Noorikuhani_2023} and radial-evolution corrections \citep{Bonvin:2013ogt,Beutler_2020,Paul:2022xfx,Addis_2025}. 

To overcome this issue, a promising approach is to expand both fields in spherical Fourier-Bessel series, eventually dealing with a spherical power spectrum \(\smash{C^{XY}_l(k,k')}\), with now \(l\) representing an angular multipole and \(k=|\bm k|\) and \(k'\) being Fourier wavenumbers \citep{2005PhRvD..72b3516C,2012A&A...540A.115R,2013PhRvD..88b3502Y}. This solution, however, is extremely computationally cumbersome \citep[see e.g.][]{2021PhRvD.104l3548G}, which is why it is more usual to expand only the angle between the lines of sight \(\hat{\bm r}=\bm r/r\) and \(\hat{\bm r}'\) in multipole series, keeping the configuration space positions in terms of the redshift corresponding to the radial comoving distances to the two perturbations, \(z\) and \(z'\). Thus, we deal with the harmonic-space power spectrum, \(C^{XY}_l(z,z')\), or its binned (a.k.a.\ tomographic) version, \(C^{XY}_{ij,l}\), where now \(i,j=1\ldots N_z\) label the \(N_z\) redshift bins which the original fields have been sliced into.

For a Gaussian random field---an ansatz that we believe is satisfied by the primordial gravitational potential---the mean and the variance are all that is needed to describe it. However, gravity is non-linear and density fluctuations that grew on top of those initial perturbations in the gravitational field to give rise to the observed LSS can no longer be described by Gaussian statistics. For this reason, we can also use higher-order correlators to study the LSS. Due to the increasing complexity of the problem, the most studied of such \(N\)-pt correlation functions is the 3-pt one, \(\zeta_{XYZ}(\bm r_1,\bm r_2,\bm r_3)\coloneqq\langle X(\bm r_1)\,Y(\bm r_2)\,Z(\bm r_3)\rangle\), where now the three comoving positions form the vertices of a triangle. The corresponding poly-spectrum is the bispectrum, \(\langle X(\bm k_1)\,Y(\bm k_2)\,Z(\bm k_3)\rangle\eqqcolon(2\,\pi)^3\,\dd(\bm k_1+\bm k_2+\bm k_3)\,B_{XYZ}(\bm k_1,\bm k_2)\), with the Dirac-delta ensuring momentum conservation in Fourier space.

As a last remark on nomenclature, we shall refer to auto- or cross- poly-spectra/\(N\)-pt correlation functions depending, respectively, on whether the \(N\) fields are all the same or not. Furthermore, if they are all analysed jointly, we shall refer to this as `multi-tracing'.

\section{Detecting relativistic effects in cosmological observables}\label{sec:rel_effects}
This section is dedicated to prospects of detecting GR effects in the cosmological observables that will be measured by the SKAO. Any of such detections, which have drawn significant interest in the community over the past decade but have not yet been achieved, due to the elusiveness of relativistic effects, will serve as a further test of GR in the ultra-weak field regime. Since GR effects contribute to shear only at second order, as discussed in \cref{sec:shear}, it is then natural to focus on the clustering of tracers as a means to detect them.

\subsection{2-pt correlations of galaxy number counts and intensity mapping}
Cosmological information can be extracted from fluctuations in the number counts by measuring the 2-pt correlation function, $\xi_{tt'}\equiv\xi_{\Delta_t\Delta_{t'}}$ presented in \cref{sec:correlators}. However, GR contributions to the observed \(\Delta_t\) are strongest on the largest cosmic scales. The scarcity of independent modes for measurements of correlation functions on such scales---also known as cosmic variance---has hitherto hindered attempts at detecting GR effects. In this context, both the SKAO continuum galaxy survey \citep{Asorey01.2026.SKA} and the SKAO \hi\ intensity mapping surveys \citep{Wolz01.2026.SKA}, reaching up to the end of re-ionisation and deeper, will have access to enormous cosmological volumes, thus allowing us to access more modes even on those largest scales. Complementary to this, the SKAO \hi\ galaxy survey \citep{Nasirudin01.2026.SKA} will deliver a dense, low-redshift catalogue of sources, also suitable for studying relativistic effects thanks to the Doppler signal scaling with the ratio of the line-of-sight velocity to the Hubble expansion rate. This happens because, at low redshift, galaxy peculiar velocities are significant while the Hubble expansion rate is smaller, boosting the Doppler imprint in the observed clustering and thus making \hi\ galaxies a promising tracer for these kinds of studies.

\begin{figure}
    \centering
\includegraphics[width=\textwidth]{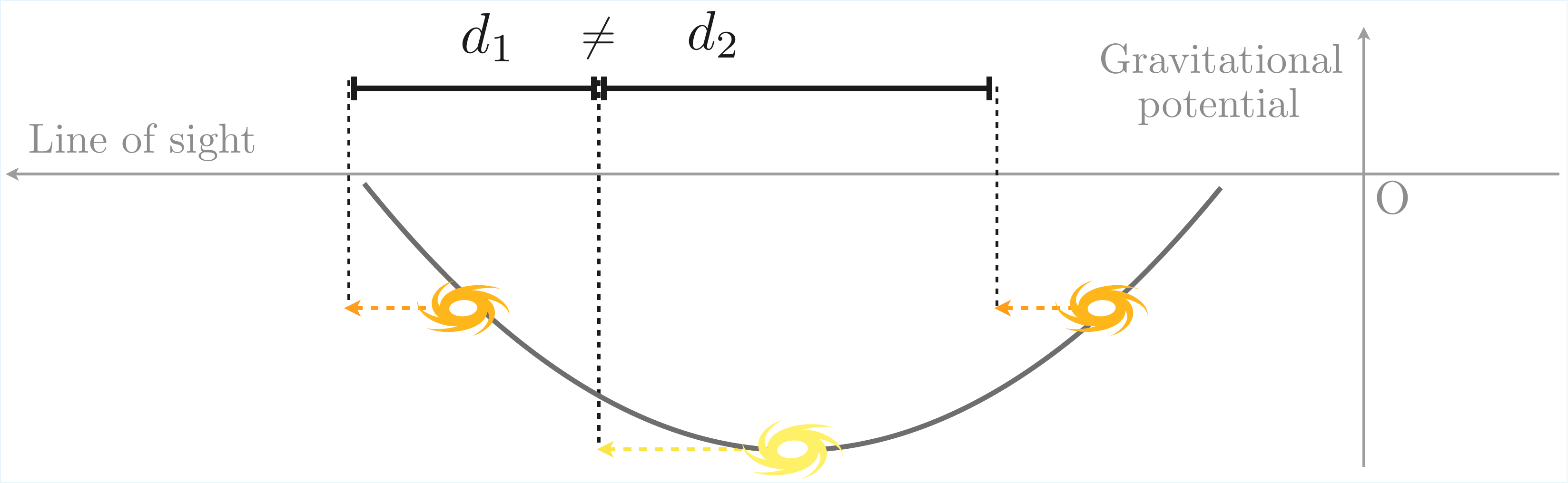}
    \caption{Gravitational redshift breaks the symmetry of 2-pt correlations, here illustrated by comparing three galaxies located in a gravitational potential. The brightest galaxy at the bottom of the potential is affected by the largest gravitational redshift, as its photons have to climb out of the deepest potential well to reach the observer at `O' (middle right). The other two fainter galaxies are located at the same height in the potential and hence experience the same amount of gravitational redshift. Thus, in redshift space, the galaxies are shifted such that the central galaxy appears to be closer to the fainter galaxy behind than to the one in front. (Illustration adapted from \citealt{Bonvin:2013ogt}.)
    }
    \label{fig:grav_red_symm_breaking}
\end{figure}

Another approach is to expand  \(\xi_{tt'}\) in Legendre multipoles about the angle \(\hat{\bm s}\) separating the two lines of sight. Standard contributions to $\Delta_t$, consisting of the density term and the RSD effect (see Eq.\ \ref{eq:Delta}), generate a monopole, a quadrupole, and a hexadecapole. Relativistic contributions to these multipoles are highly subdominant \citep[e.g.][]{Jelic-Cizmek:2020pkh}, but GR effects can also induce a dipole in the 2-pt correlation function and power spectrum \citep{McDonald_2009,Bonvin:2013ogt}. However, for this dipole not to vanish it is essential to consider the cross-correlation between two different tracers, \(t\ne t'\), with distinct properties. This can be achieved by cross-correlating any of the aforementioned radio tracers with others at different wavelengths or, in the case of \hi\ and continuum galaxies, by splitting a parent sample into faint (\(t={\rm F}\)) and bright (\(t'={\rm B}\)) sub-samples. \Cref{fig:grav_red_symm_breaking} illustrates how such a split may create an asymmetry and thus a dipole in the galaxy correlation function.

\begin{figure}
    \centering
    \includegraphics[width=0.60\linewidth]{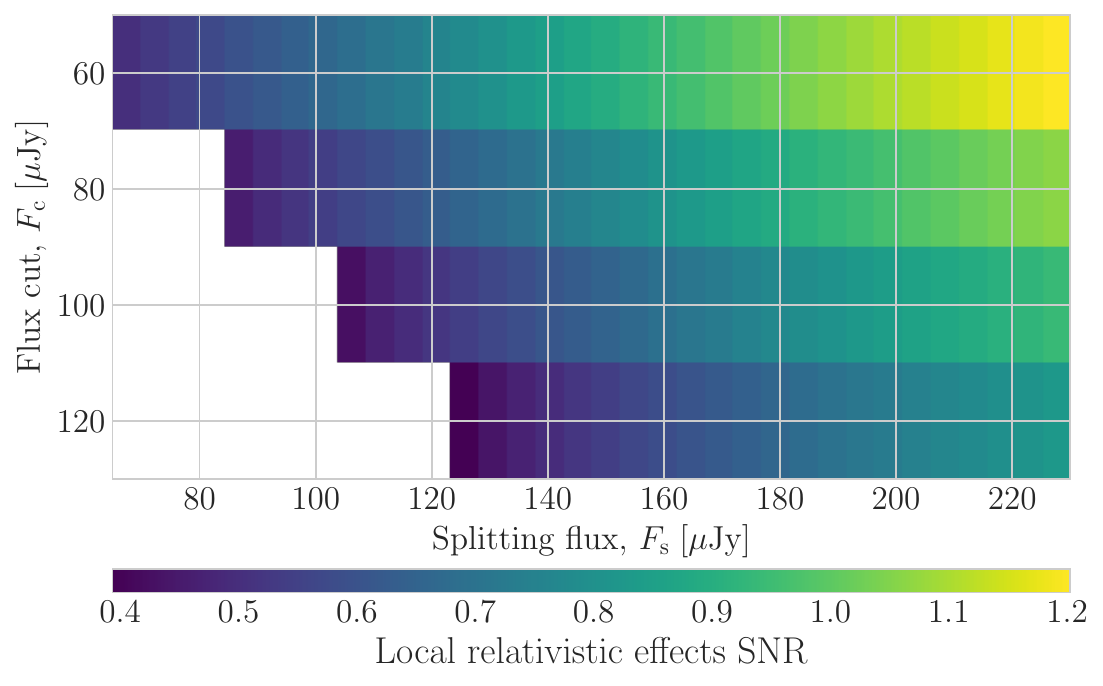}
    \includegraphics[width=0.375\linewidth]{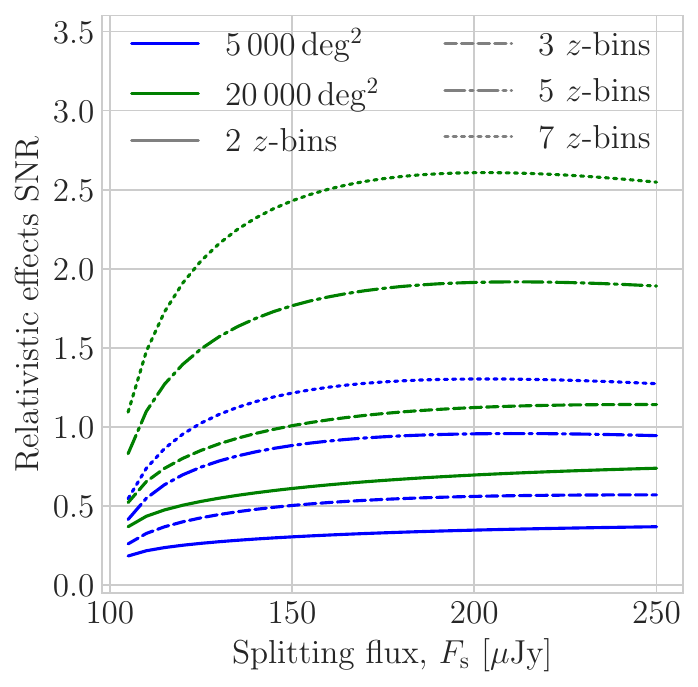}
    \caption{Cumulative SNR of the relativistic contributions in a faint-bright \hi\ cross-spectrum, over the interval $0.1\lesssim z \lesssim 0.6$. \textit{Left:} In a $P_{\rm FB}(\bm k)$ measurement, the SNR represents the statistical distance between data, affected by local GR effects, and a simple model which does not include relativistic corrections. 
    \textit{Right:} SNR, estimated via a null-hypothesis of $\Delta_t^{\rm (Dop)}=\Delta_t^{\rm(rel)}=0$, in the case of a multi-tracer faint-bright angular power spectrum analysis. We assume a $\qty{100}{\micro\Jy}$ flux cut, together with both a conservative sky coverage of $5\,000\, \deg^2$ (blue curves) and an optimistic one of $20\,000\,\deg^2$ (green curves). Results in both panels illustrate how with an optimal choice of $F_{\rm s}$ we can boost the GR signal detection significance.}
    \label{fig:Doppler_X2}
\end{figure}

\paragraph{Faint-bright cross-spectrum.}
We can subdivide the \hi\ galaxy population into a faint and a bright sub-sample, and then measure cross-correlations between them. 
Following the procedure outlined in \citet{Montano2024_1,Montano2024_2}, the left panel of \cref{fig:Doppler_X2} quantifies the relevance of the local relativistic corrections---whose main contribution comes from the Doppler term---in a measurement of the Fourier-space faint-bright cross-spectrum, $P_{\rm FB}(\bm k)$, across the redshift range $z\in[0.1,\,0.6]$. It assumes a surveyed area of $5\,000\, \deg^2$ \citep[see][]{SKA:2018ckk}, limited by the observation time, although SKAO's sky visibility could allow for a wider, more optimistic, coverage. The signal-to-noise ratio, SNR, depends on both the critical flux of the survey, $F_{\rm c}$, and the splitting flux between the two populations, $F_{\rm s}$. We can thus fix $F_{\rm s}$ in order to maximise the relativistic signal. 
The scaling of the detection significance versus the observed sky fraction, \(\snr(f_{\rm sky})=\sqrt{f_{\rm sky}}\,\snr(1)\), shows that a \(15\,000\) to \(20\,000\,\deg^2\) \hi\ galaxy survey with AA4 specifications will reach a $\rm{SNR}\!\sim \!2$ with a measurement of the faint-bright cross-power spectrum alone. This, in turn, may lead to a robust detection of the relativistic Doppler effect by adding the constraining capability of auto-correlation Fourier-space spectra \citep[following e.g.][]{Montano2024_2}.

\paragraph{Faint-bright multi-tracing.}
The two sub-samples can be seen as independent realisations of the same underlying matter density field, thus allowing for multi-tracer analyses \citep{Seljak:2008xr,McDonald&Seljak2009}. Such a joint analysis of the auto-spectra $P_{\rm FF}$ and $P_{\rm BB}$, and the cross-spectrum $P_{\rm FB}$, will be able to limit cosmic variance and  probe GR signatures with further precision \citep[e.g.][]{White_2009,Abramo_2013,Alonso:2015sfa,Fonseca:2015laa,Abramo_2022}. 
Thus, the right panel of \cref{fig:Doppler_X2} displays the statistical significance of $\Delta_t^{\rm (Dop)}\subset\Delta_t^{\rm (vel)}$ and $\Delta_t^{\rm (rel)}$, when probed by a multi-tracer faint-bright power spectrum, now considered in harmonic space.
It essentially extends the outcomes of \citet{Novara2025_preprint} to the SKAO \hi\ galaxy survey, showing that an optimal definition of the redshift bins will be able to counteract the loss of information due to the projection on spherical shells.
Alongside, the flexibility in the binning choices may also be exploited to consider radio continuum galaxies \citep[see][]{Asorey01.2026.SKA}. Although we found it unlikely to achieve a detection of relativistic Doppler with continuum galaxies, they may increase the constraining power of other targets, if cross-correlated with them.

\begin{figure}
    \centering
\includegraphics[width=0.75\textwidth]{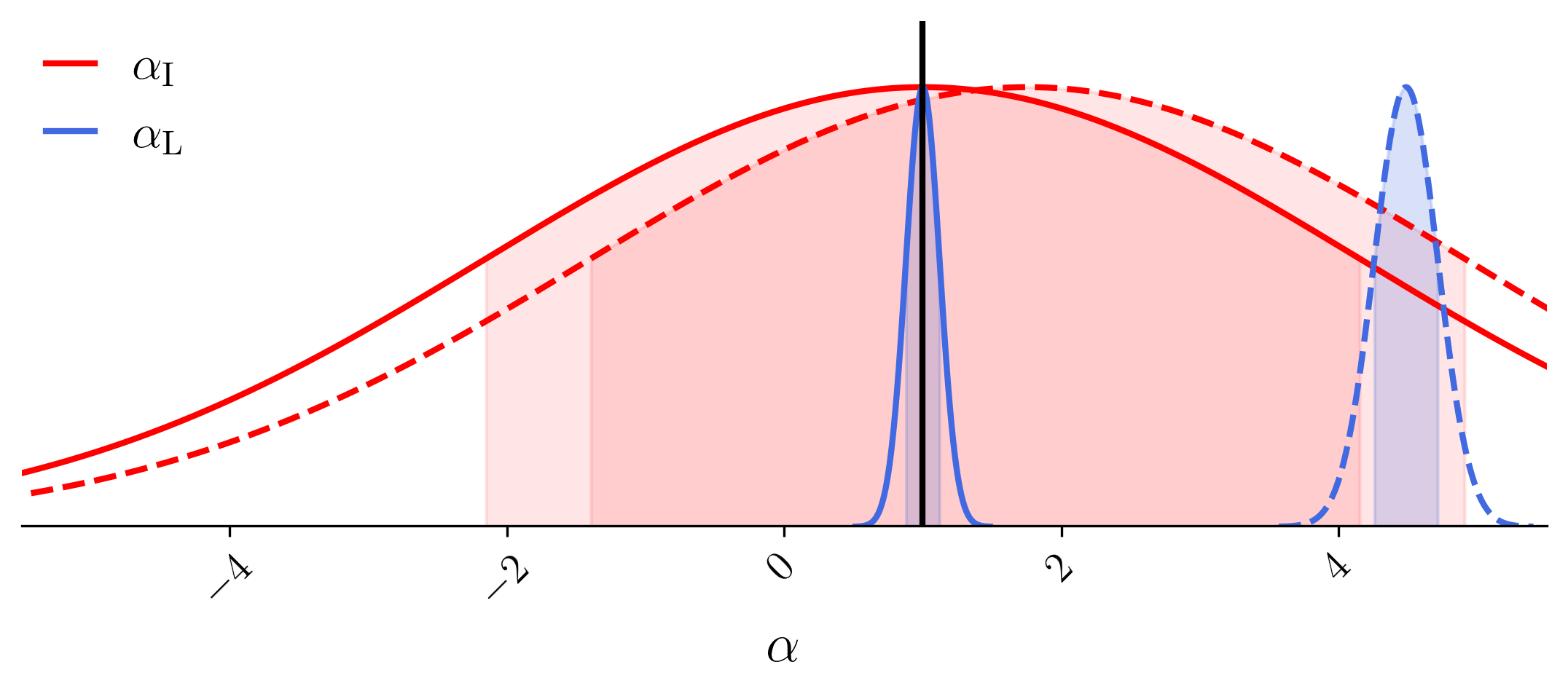}
    \caption{Forecast constraints on the amplitude of local and integrated relativistic contributions, $\alpha_{\rm L}$ and $\alpha_{\rm I}$, in a multi-tracer analysis for the Legendre multipoles of the power spectrum (up to the hexadecapole), having included (solid lines) or not (dashed lines) wide-separation corrections in the theoretical model. We adopt a flux cut of $F_{\rm c} = \qty{100}{\micro\Jy}$ and a $F_{\rm s} = \qty{150}{\micro\Jy}$ splitting flux. The probability distributions are normalised to equal peak height, with the vertical black solid line marking the ground truth, $\alpha_{\rm L}=\alpha_{\rm I}=1$.}
    \label{fig:multipole_constraints}
\end{figure}

Alternatively, one can consider the same type of faint-bright split approach but instead using the multipoles of the Fourier power spectrum, where one would use the so-called Yamamoto estimator \citep{Yamamoto_2005,Scoccimarro_2015,2022JCAP...01..061C}. \Cref{fig:multipole_constraints} shows the constraints on the amplitude $\alpha_{\rm L}$ of the local relativistic terms and that of the integrated ones, $\alpha_{\rm I}$, in the scenario where one neglects wide-separation terms (dashed). As such, there is a shift in the best fit measurement of the amplitudes for a full multi-tracer analysis, as described in \citet{Addis_2025_new}. For this low redshift survey, the bias from wide-separation corrections is significant, amounting to as much as $\approx 15\,\sigma$. When we include wide-separation corrections, we also gain additional relativistic terms in the form of the wide-separation corrections to the relativistic contribution. Thus, we forecast a precise measurement of the local effects, $\sigma(\alpha_{\rm L})=0.12$, although the integrated effects themselves are unlikely to be detectable for this sample, $\sigma(\alpha_{\rm I})=3.2$. Note that here no uncertainty in the bias modelling is assumed. 

\paragraph{Radio-optical cross-spectrum.}
The variety of LSS tracers provided by the SKAO allows for cross-correlations with traditional optical/near-infrared cosmological surveys over a broad swathe of redshifts. At high-\(z\), \hi\ intensity mapping can be analysed in cross-correlation with spectroscopic galaxy catalogues like \textit{Euclid}'s or the DESI emission-line galaxy sample, with the advantage of negligible noise in the cross-spectrum. At lower redshift, we can exploit the almost perfect overlap between the high density of \hi\ galaxies and the DESI bright galaxy sample.

Detecting the statistical effect of peculiar velocities would be crucial to understand how these are related to the gravitational potential $\Psi$, i.e.\ testing the Euler equation cosmologically. One can construct an anti-symmetric estimator \citep{Bonvin:2018ckp,2021JCAP...12..003F} on very large-scales that would be sensitive to the effect of peculiar velocities, i.e.\ the Doppler term, in the observed number counts of tracers of the LSS. The authors find that synergies between a catalogue of \hi\ galaxies built from an AA$^\ast$ configuration and DESI can identify the Doppler term up to $3\,\sigma$, whilst only upgrades to the SKAO can deliver game-changing observations with radio observables alone in auto-correlation.

\begin{figure}
\centering
\includegraphics[width=0.32\linewidth]{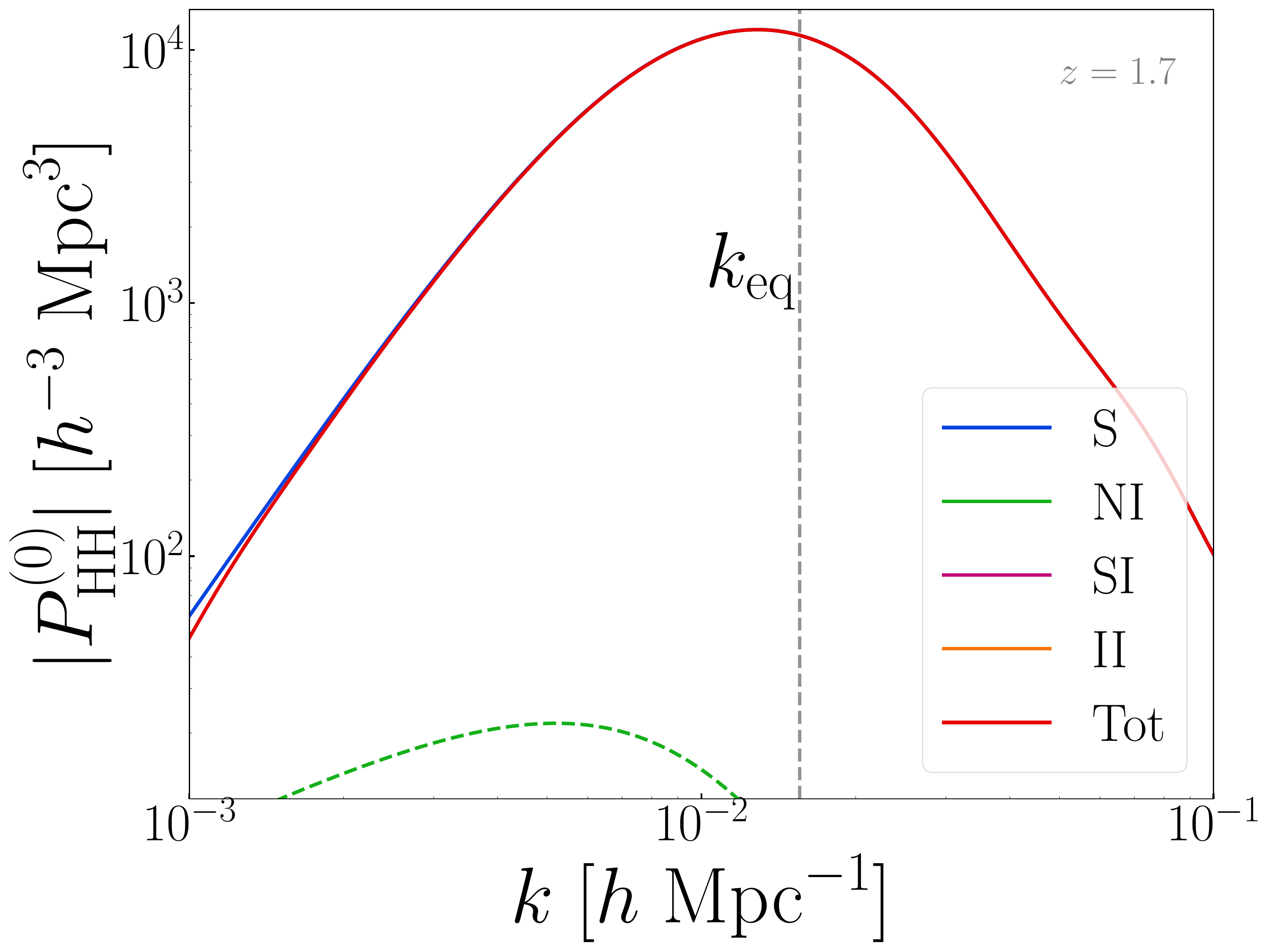} \,
\includegraphics[width=0.32\linewidth]{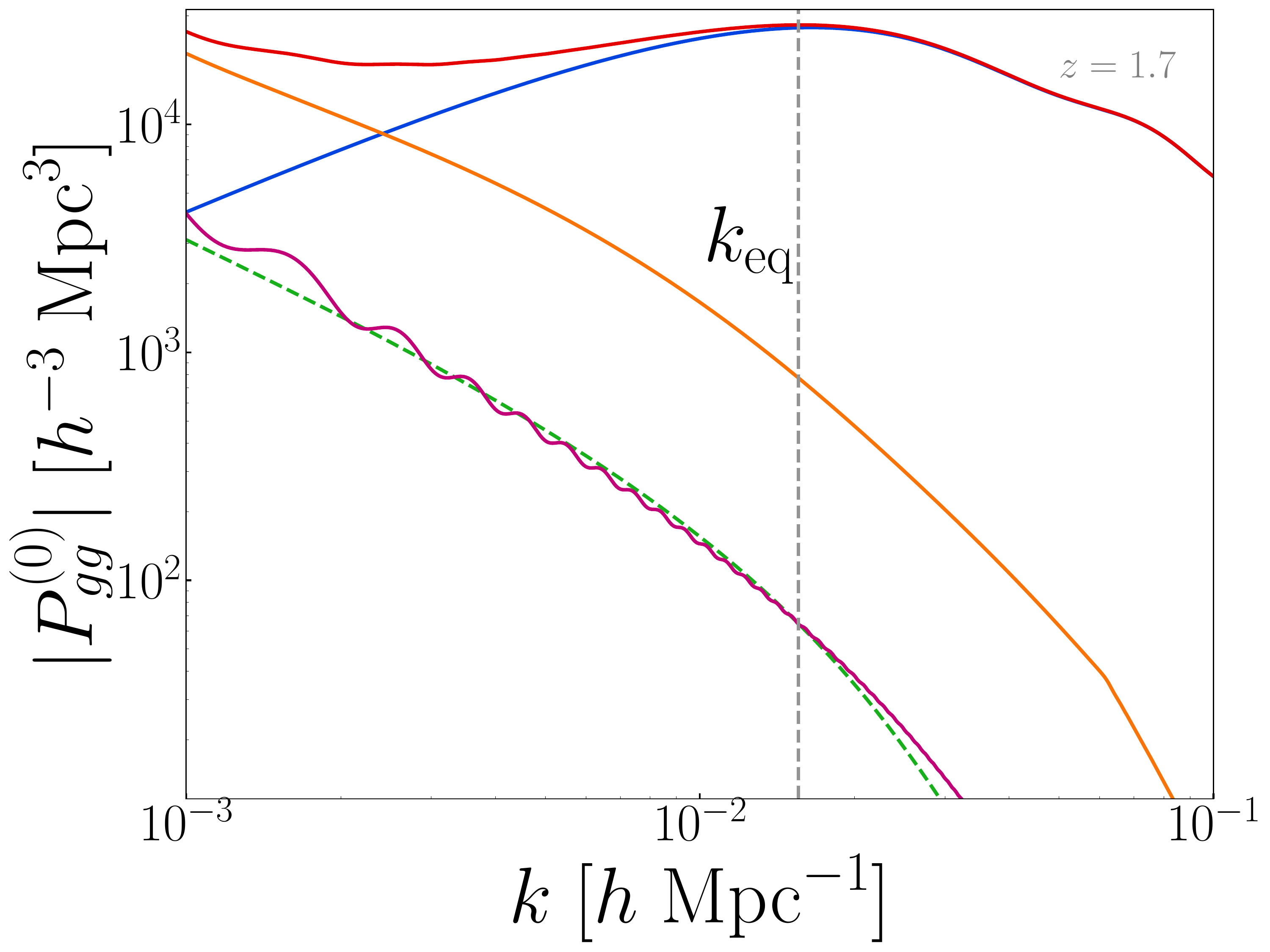} \,
\includegraphics[width=0.32\linewidth]{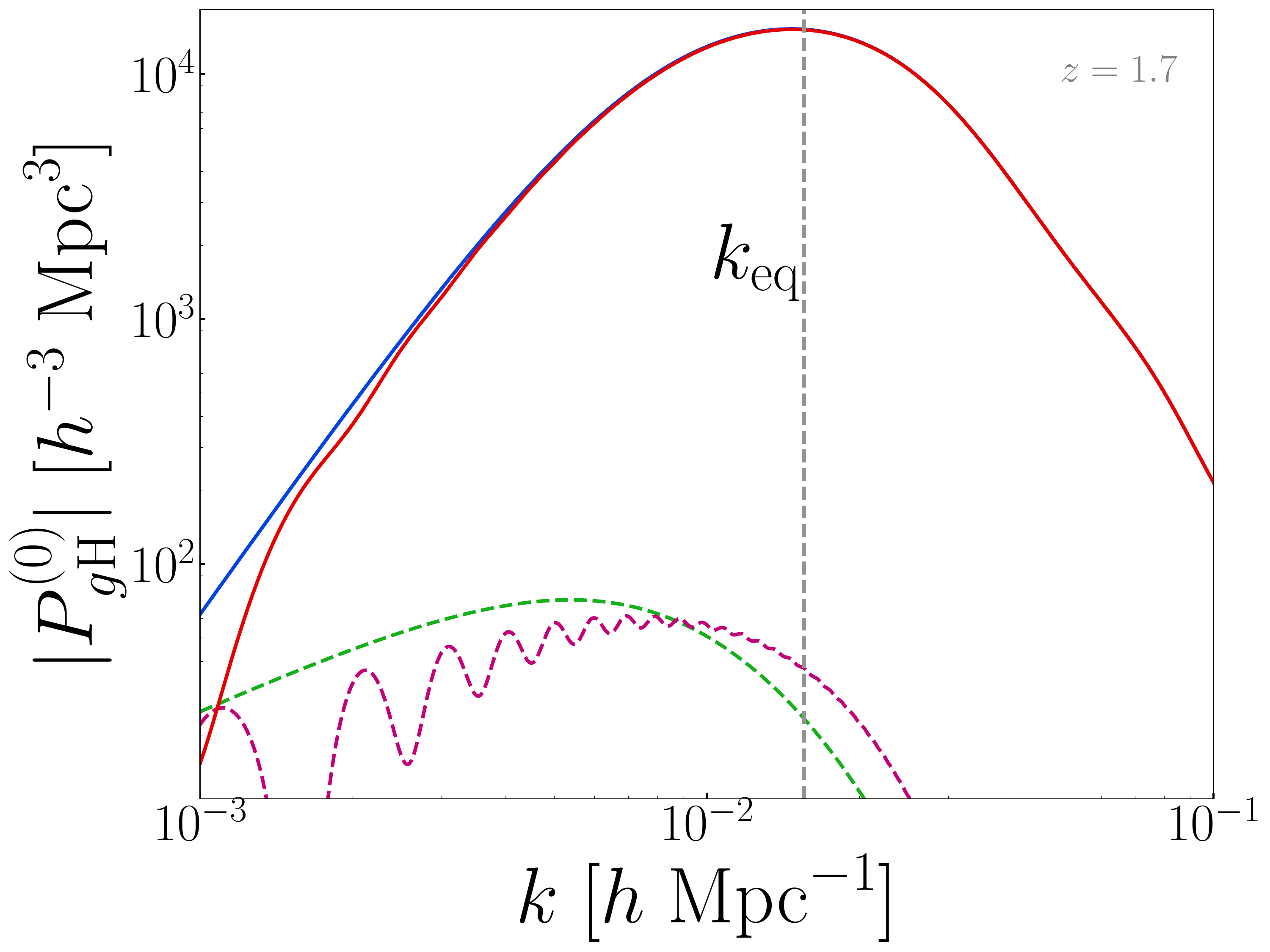} 
\caption{Legendre monopole of the SKAO \hi-\hi\ power spectrum (left panel), \textit{Euclid} galaxy-galaxy power spectrum (middle panel), and their cross-spectrum (right panel) at $z=1.7$. The plots display the importance of the non-integrated (NI), integrated (I), and total NI+I (`Tot') contributions, where \(\mathrm{I} = \mathrm{SI} + \mathrm{II}\).}
\label{fig:monopolesHIgal}
\end{figure}

Regarding radio-optical synergies at high redshift, instead, we consider the SKA-Mid Band 1 intensity mapping survey and the \textit{Euclid} H$\alpha$ spectroscopic galaxy survey, both probing the LSS of the Universe in the redshift range \(z\in[0.9,1.8]\) and with an estimated overlapping sky area of approximately \(8\,000 \,\deg^2\).
\Cref{fig:monopolesHIgal}, showing the Legendre monopole of the different power spectra at $z=1.7$, demonstrates that relativistic integrated contributions (`SI', purple curves, and `II', orange curves) are negligible for the \hi\ auto-spectrum (left panel) and for the \hi-galaxy cross-spectrum (middle panel), owing to the near cancellation of the lensing term in brightness temperature fluctuations. In contrast, in the \textit{Euclid} galaxy auto-spectrum (right panel), the non-integrated corrections (`NI', green curves) become relevant at intermediate scales ($\sim10^{-2}\,h\,\si{\per\mega\parsec}$), while the integrated-integrated component dominates at ultra-large scales ($\sim10^{-3}\,h\,\si{\per\mega\parsec}$). This implies that the modelling of the \hi\ power spectrum, already complicated on the largest scales because of the impact of foreground removal and signal loss \citep[see e.g.][]{2023MNRAS.523.2453C}, as well as by imprints of inflationary physics \citep{2020MNRAS.499.4054C}, is not expected to be affected by neglecting a precise modelling of relativistic effects.
   

\subsection{Bispectrum of galaxy number counts and  intensity mapping}
The dominant relativistic signature in galaxy clustering, the Doppler correction, can be best exposed in the 2-pt correlation function and power spectrum only when considering cross-correlations. The bispectrum, on the other hand, provides a powerful complementary approach: a `smoking gun' imaginary dipole signal is accessible in the bispectrum of a single tracer,  without the need for cross-correlations \citep{Clarkson:2018dwn,Jeong:2019igb}. This also applies to \hi\ intensity mapping \citep{2021JCAP...06..039J}. Higher-order statistics such as the 4-pt correlation function also contain these signatures \citep{Paul:2024uim}, extending the treatment of relativistic effects in the non-linear regime, generalising \cref{eq:Delta}. 

\begin{figure}
    \centering
\includegraphics[width=0.75
\textwidth]{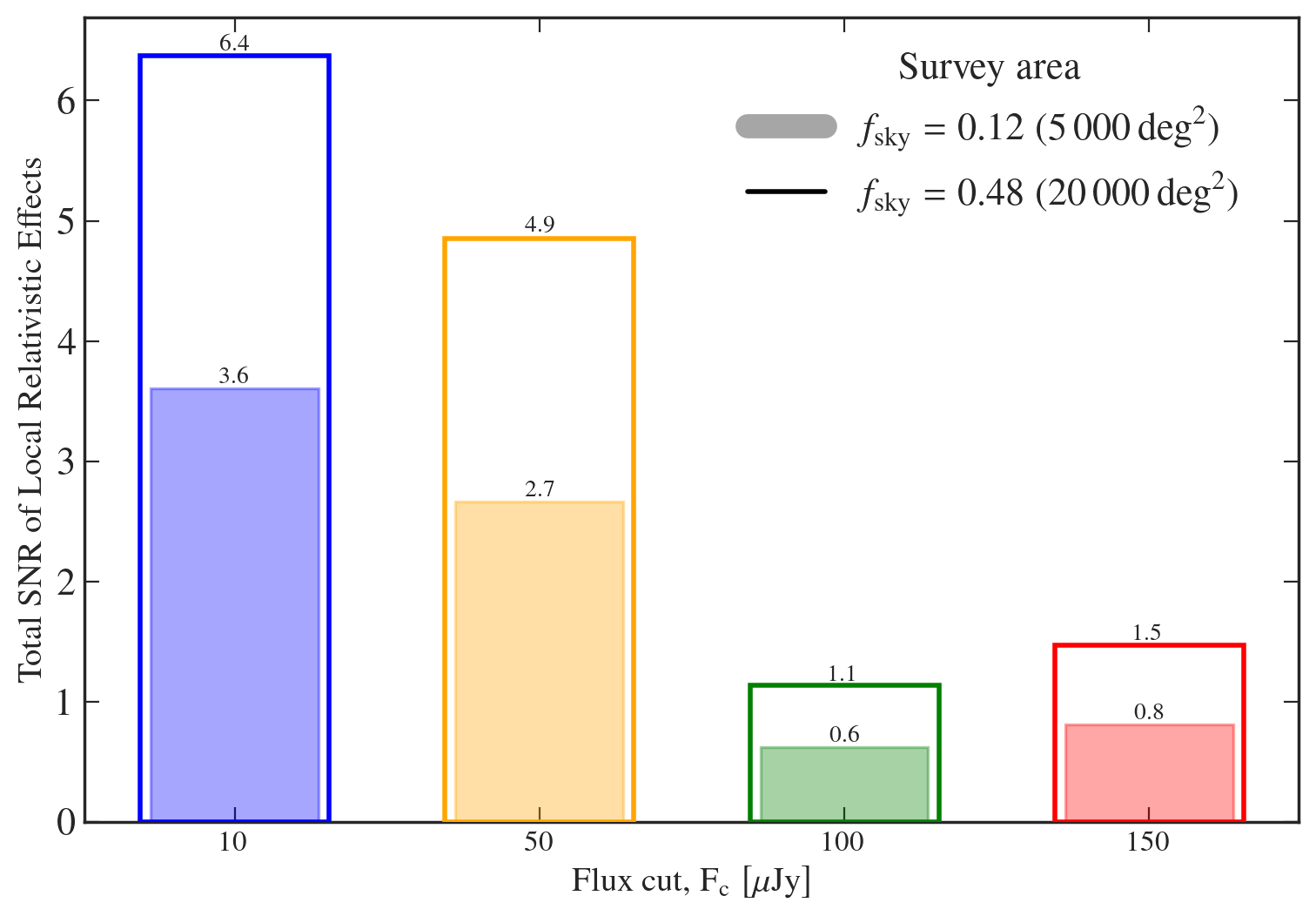}
    \caption{The forecasted cumulative signal-to-noise ratio of the relativistic contributions to the galaxy bispectrum for SKAO \hi\ galaxies, assessed at various flux cuts. For more realistic flux cuts, $\qtylist{150;100}{\micro\Jy}$, the signal is unlikely to be detectable. Future improvements to the flux sensitivity of the instrument could prove fruitful in this regard.}
    \label{fig:bispectrum_SNR}
\end{figure}

However, detecting such contributions using the auto-bispectrum remains inherently challenging as we fight against the limited number of large-scale modes and noise in this range. \Cref{fig:bispectrum_SNR} presents forecasts for the cumulative signal-to-noise ratio associated with the general relativistic contributions to the galaxy bispectrum for SKAO \hi\ galaxies. We assess the expected sensitivity achievable with the single-tracer bispectrum and establish a baseline for future improvements aimed at maximising scientific return of SKAO on ultra-large scales. In fact, as we see in \cref{fig:bispectrum_SNR}, the scientific return can be optimised by varying the flux cut of the sample. Whereas lower flux cuts generally improve the signal-to-noise ratio by increasing the number density, the behaviour at high flux cuts (\(F_{\rm c} \approx \qtyrange{150}{100}{\micro\Jy}\)) is non-monotonic: the bispectrum covariance is shot-noise dominated, thus the noise penalty saturates. In this regime, enhancements from magnification and evolution bias---particularly at intermediate to high redshifts---outweigh the additional noise contribution. A full multi-tracer bispectrum can exploit this further \citep{2026arXiv260300244R}.  These results mark an initial step in exploiting the full potential of the clustering of \hi\ galaxies to test gravity. 

\begin{figure}
    \centering
    \includegraphics[width=0.4\textwidth]{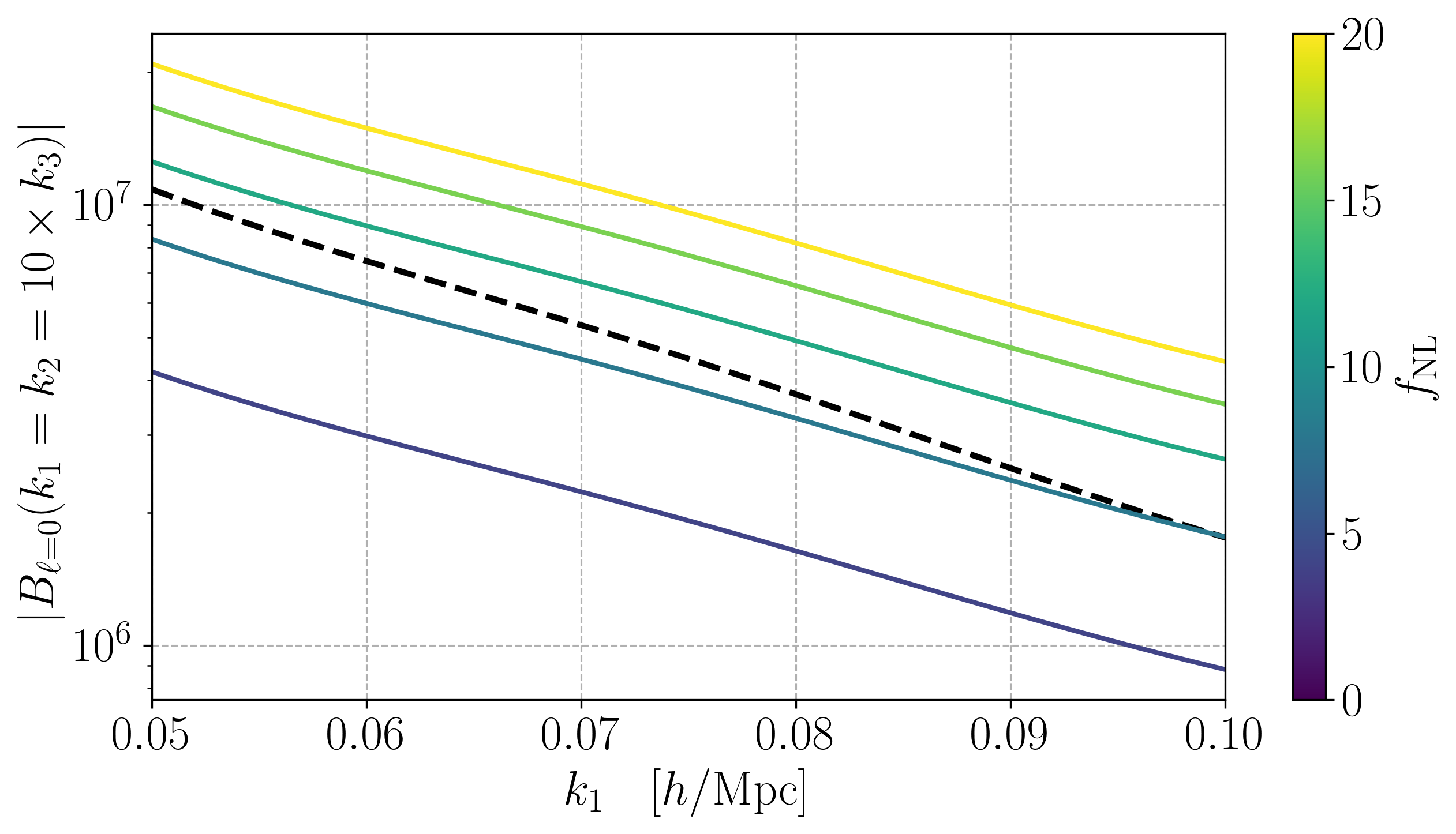}
    \includegraphics[width=0.59\linewidth]{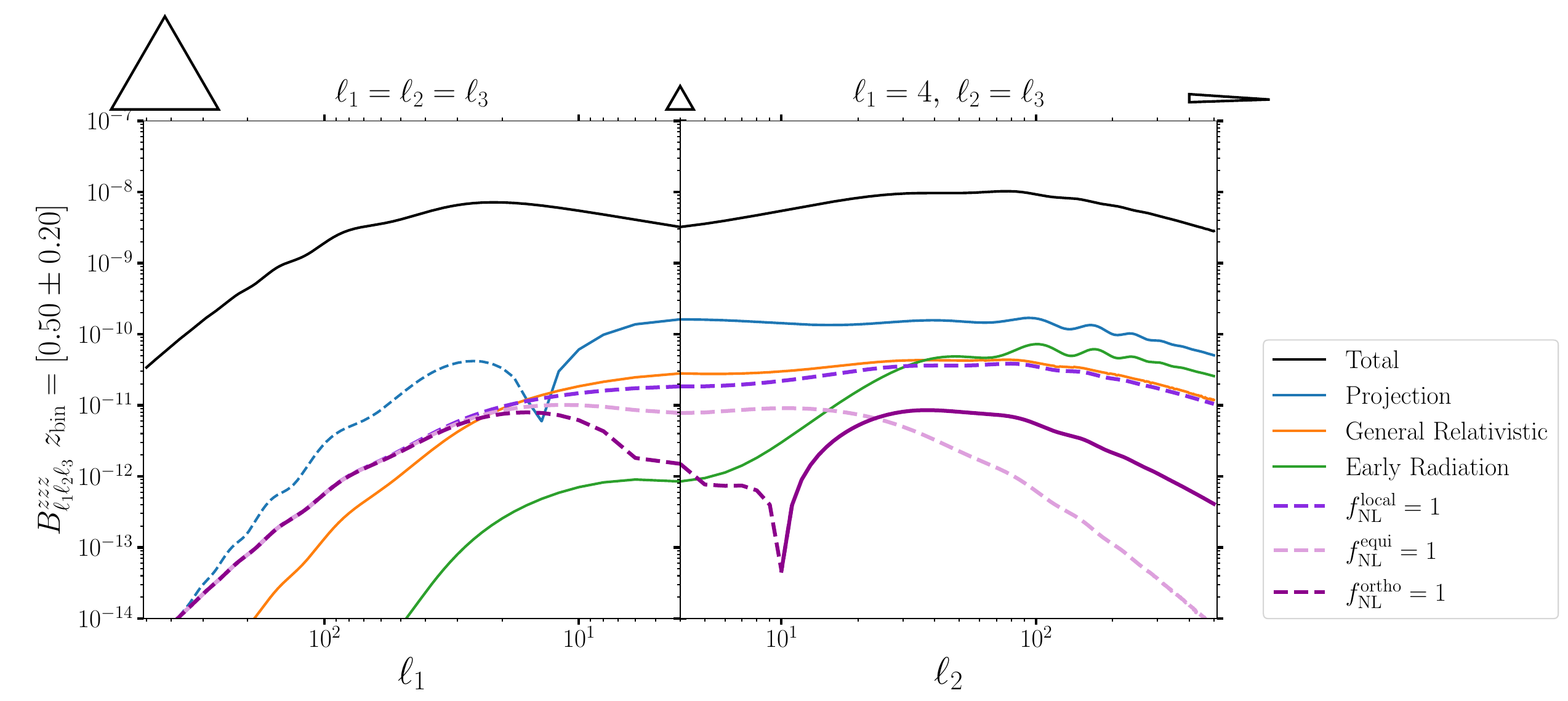}
    \caption{Bispectrum of SKAO \hi\ galaxies including both GR contributions and PNG, in both Fourier and harmonic space (left and right panel, respectively). Coloured lines in the right panel refer to different values of the PNG amplitude parameter, \(f_{\rm NL}\), with the GR prediction being the black, dashed curve. In the left panel, the total GR bispectrum (solid black) is compared to the Newtonian approximation (dotted black), with individual relativistic contributions shown: relativistic projection effects (blue), GR corrections (orange), and early-time radiation effects (green). Long-dashed purple lines show the PNG signals for local (dark), equilateral (light), and orthogonal (medium) shapes with $f_{\rm NL}=1$. Triangle diagrams at the top illustrate the respective configurations. Solid and short-dashed curves respectively refer to positive or negative bispectra.}
    \label{fig:angular_bispectrum}
\end{figure}

Moreover, an important motivation for studying the bispectrum is its role 
in detecting primordial non-Gaussianity (PNG) via the scale-dependent bias that local PNG induces \citep[see also][]{Fonseca01.2026.SKA}. 
On ultra-large scales, however, relativistic terms generate scale-dependent signatures that closely resemble those sourced by local PNG \citep{Bruni:2011ta,2015MNRAS.451L..80C,Maartens:2020jzf}. If unaccounted for, these relativistic contributions can bias measurements of the local PNG amplitude, $f_{\rm NL}$ \citep{Umeh_2017,DiDio:2016gpd,2024arXiv240706301R,Addis_2025}.
To illustrate this interplay between relativistic effects and PNG in the bispectrum, we present in \cref{fig:angular_bispectrum} the bispectrum of SKAO \hi\ galaxies, both in Fourier space (left panel) and in harmonic space (right panel). For the former, we show the local-type PNG signal in a squeezed limit of the Legendre monopole of the bispectrum for different values of the PNG amplitude parameter $f_{\rm NL}$, in comparison to the local relativistic signal (black dashed), assuming a nominal $\qty{100}{\micro\Jy}$ flux cut. Clearly, the scale dependence of the PNG signal closely resembles that of GR effects. 

Similarly, in the right panel we see the harmonic-space bispectrum, $B^{zzz}_{\ell_1\ell_2\ell_3}$, again for an SKAO \hi\ galaxy survey in a redshift bin centred on $0.50$ and of width $0.20$.\footnote{The angular bispectrum of galaxy count is evaluated using the code \href{https://github.com/TomaMTD/ang_bispec}{\texttt{ang\_bispec}} \citep{Montandon:2025uul}.} In this framework, all types of local and integrated GR terms can be easily introduced \citep{DiDio:2014lka, DiDio:2015bua, DiDio:2018unb, Assassi:2017lea, Montandon:2022ulz, Montandon:2025uul}. Here, we can compare the full GR prediction with its Newtonian approximation across two configurations: equilateral (\(\ell_1=\ell_2=\ell_3\)) in the left panel (with reverted abscissas) and squeezed (\(\ell_1=4\), \(\ell_2=\ell_3\)) in the right panel. Solid(dashed) curves represent positive(negative) bispectra. The relativistic projection effects, also sourcing the bispectrum dipole, peak at large scales in the equilateral configuration, where they represent \(10\%\) of the signal. In the squeezed limit, they constitute the dominant relativistic contribution. 
Crucially, the relativistic signatures exhibit amplitude and angular scale dependence similar to that of PNG, highlighting the degeneracy challenge: neglecting relativistic
effects when constraining $f_{\rm NL}$ could lead to significant biases.

However, this also presents an opportunity: by modelling both contributions simultaneously, the bispectrum offers a means to constrain inflationary physics and test GR on cosmological scales. Thus, the relativistic bispectrum  provides an additional avenue to test gravity on cosmological scales in addition to  the power spectrum and 2-pt correlation function, with the added benefit of directly mitigating biases in the constraints on inflationary models.

\section{Detecting signatures of modified gravity}\label{sec:mg}

The idea behind modified gravity theories is to alter the gravitational dynamics on large scales, comparable to or greater than the Hubble radius, whilst preserving the successful predictions of GR on Solar System and smaller scales. This can be achieved through several mechanisms. The best-studied examples are scalar-tensor theories, such as Horndeski, beyond-Horndeski, and DHOST models, which introduce one additional scalar degree of freedom coupled to the metric and modify both the propagation of gravitational waves and the effective gravitational constant \citep{Horndeski:1974wa,Deffayet:2011gz,Langlois:2015cwa}.

However, other field contents are also viable. Tensor-tensor theories (e.g.\ bimetric gravity) introduce an additional dynamical metric, leading to massive graviton degrees of freedom \citep{Hassan:2011hr}. Vector-tensor theories, such as generalized Proca or Einstein-\AE ther models, introduce vector fields that can drive the acceleration or modify the growth of structure \citep{Heisenberg:2014rta,Jacobson:2007veq}. Non-metric theories, including teleparallel gravity or symmetric teleparallel $f(Q)$ models \citep{BeltranJimenez:2019esp}, reformulate gravitation in terms of torsion or non-metricity instead of curvature, offering geometrically distinct routes to cosmic acceleration.

To remain consistent with local and astrophysical tests, many of these theories rely on screening mechanisms (chameleon, Vainshtein, or symmetron effects) that suppress deviations from GR in high-density environments \citep[see][]{Khoury:2003aq,Babichev:2013usa}. The rich phenomenology of these frameworks, ranging from modified GW speed to scale- and/or time-dependent effective couplings, opens multiple observational windows for testing gravity on cosmological scales. At the cost of limiting constraints to insights on specific models, \hi\ intensity mapping over large SKAO-like volumes is, for example, expected to be sensitive to the shape of scalar potentials and effective couplings, as has been shown for redshift $z>6$ \citep{2020JCAP...05..038L}.

In addition to exploring specific models, an important complementary strategy involves model-independent parametrisations of cosmic acceleration. The effective field theory (EFT) of dark energy and modified gravity provides a unified framework that captures the low-energy dynamics of a wide class of single scalar field modified gravity and dark energy models in terms of a finite set of time-dependent functions appearing in the perturbed action \citep{Frusciante:2019xia}. This approach allows for a systematic comparison between theory and data without committing to a specific Lagrangian.

\begin{figure}
    \centering
\includegraphics[width=0.6\textwidth]{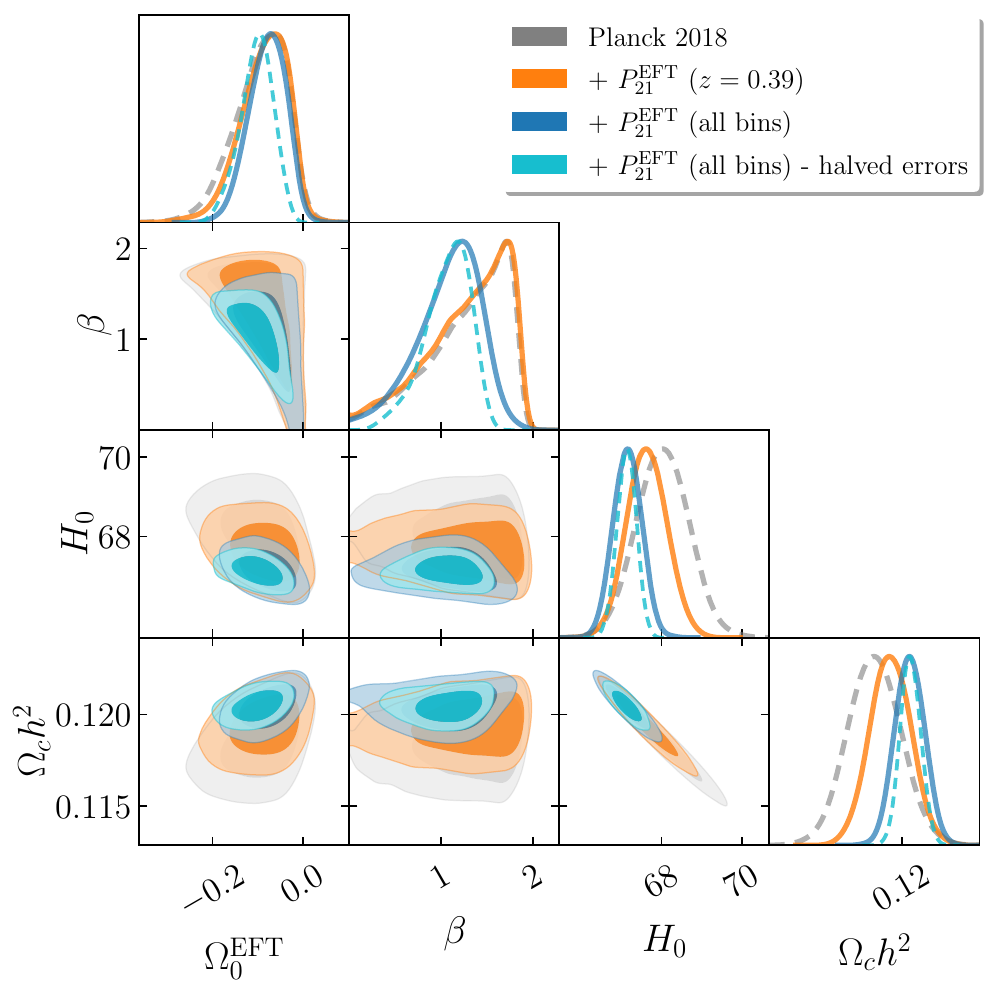}
    \caption{Results from \cite{Berti:2021ccw}. Forecast joint constraints on parameters of EFT models of dark energy from a MeerKLASS-like \hi\ intensity mapping survey in combination with \textit{Planck} 2018 data. We compare the constraining power of a single bin \hi\ auto-spectrum detection $(z=0.39)$ with a five-bin one, at $z=0,\,0.39,\,0.53,\,0.67,\,2.5$ (`all bins' in legend). We also consider a more ideal case with smaller errors. The label \textit{Planck} 2018 stands for TT, TE, EE + lowE + lensing.}
    \label{fig:EFT_constraints}
\end{figure}

From an SKAO perspective, a MeerKLASS-like \hi\ intensity mapping survey would improve the constraints on the EFT functions with respect to \textit{Planck}'s current bounds \citep{Berti:2021ccw}. \Cref{fig:EFT_constraints} shows the results for an EFT model described by a running of the gravitational constant (namely, of the Planck mass) that evolves exponentially with the scale factor, i.e.\ $\mathrm{\Omega}^{\mathrm{EFT}}(a) =  \exp ( \mathrm{\Omega}_0^{\mathrm{EFT}} a^\beta ) -1$. The constraints presented are obtained by combining the forecast MeerKLASS-like data sets with \textit{Planck} 2018 observations. Whilst the power-law index $\beta$ is left almost unconstrained, we observe that \hi\ intensity mapping tomography will tighten the error bar on $\mathrm{\Omega}_0^{\mathrm{EFT}}$ by $\sim 25\%$ with respect to \textit{Planck} alone. Furthermore, if e.g.\ observation time could be extended so as to halve instrumental noise errors, the increment would go up to $\sim 35\%$.  Instead, the constraining power of single-bin observations is insufficient to improve upon \textit{Planck}'s results. 

It is also possible to constrain dark energy models with \hi\ intensity mapping observations alone, but using other experiments in different redshift intervals, provided that there is an excellent foreground removal \citep{2022JCAP...01..060W,2023SCPMA..6670413W}.
In particular, with SKA-Mid ($0.35<z<0.8$), the errors on the parameters of \lcdm---with an equation-of-state parameter for dark energy different from a cosmological constant---are lower than those obtained by current measurements from \textit{Planck} in combination with baryon acoustic oscillations and type Ia supernov\ae. This happens because  different parameter degeneracies are lifted by such probe combinations. In addition, this strategy can significantly improve constraints on the Hubble constant $H_0$ and on the cosmographic parameters $q_0$, $j_0$, and $s_0$, which are phenomenological direct tests of the deviation of dark energy from the \lcdm\ model (see also \cref{sec:additional}). Another model independent approach may involve an effective dark fluid equation of state parameter that encapsulates modified gravity theories that alter both the Universe's background expansion and the growth of the LSS, for which SKAO observations in combination with surveys such as \textit{Euclid} appear promising \citep[e.g.][]{2026arXiv260116899S}.

Phenomenological parametrisations of modified gravity, on the other hand, offer a more direct connection to observables: the modified Poisson equation and lensing potential can be expressed through the functions $\mu(a,k)$ and $\eta(a,k)$, as depicted in \cref{fig:four_fields}, which respectively describe modifications to the clustering of matter and to light deflection \citep{Zhang:2007nk, Amendola:2007rr,Pogosian:2010tj}. Deviations from GR would appear as scale- and/or redshift-dependent departures of these functions from unity, which represents their GR limit. These functions are more naturally defined in Fourier space rather than configuration space in order to preserve statistical homogeneity. This is the convention that we adopt in the following to compare with mainstream literature on the topic.\footnote{In the presence of a non-trivial scale-dependence, the conversion to configuration space leads to a less convenient form of the Poisson equation involving a convolution with a non-local derivative operator, so that the gravitational potential at one point depends on the density field at other points. There are however specific cases where a configuration-space definition is more suitable, for example in the presence of screening effects, where the modifications are non-perturbative and environment-dependent.}

Constraints on $\mu$ and $\Sigma=(1+\eta)\,\mu/2$ from SKAO \hi\ and continuum galaxy clustering and radio cosmic shear, in combination with optical/near-infrared probes, have been shown to be very competitive \citep{Casas:2022vik}, with constraints ranging from $0.83\%$ to $2.55\%$ on $\mu$ and from $0.55\%$ to $1.2\%$ on $\Sigma$. 
Similarly, \hi\ intensity mapping is able to constrain $\mu$ over a wide range in redshift (and scale), with similar percent-level constraining power also expected at redshifts higher than those of large-scale galaxy surveys  \citep{2018JCAP...10..004H}.

Lastly, another widely used diagnostic to detect deviations from GR is the growth index, $\gamma$, which parameterises the growth rate of cosmic structures as $f(a) = [\Omega_{\rm m}(a)]^\gamma$. GR with a cosmological constant---that is, \lcdm---predicts $\gamma \simeq 0.55$, whilst modified gravity theories generally predict different values or redshift dependencies \citep{Linder:2005in}. This framework allows for data-driven reconstruction of the gravitational interaction and serves as a key tool in forthcoming analyses of the LSS. \citet{2020JCAP...09..054V} proposed studying the growth index $\gamma$ by using the angular power spectrum of SKAO \hi\ galaxies, as it naturally includes the Doppler effect and gravitational lensing and does not require correcting for the Alcock-Paczy\'nski effect. Combining SKA1 with DESI or \textit{Euclid}, it is possible to reach an accuracy of $\sim2\%$ on $\gamma$ \citep[see also][for a multi-tracer approach]{2024EPJC...84...95D}. Moreover, by using the bispectrum in combination with the \hi\ intensity mapping power spectrum, sub-percent precision on \lcdm\ parameters can be attained, in an extended parameter space including $\gamma$ as well as the dark energy equation-of-state parameters $w_0$ and $w_a$ \citep{2022JCAP...11..003K}.

\section{Additional probes of gravity}\label{sec:additional}
%
%

Hitherto, we have focussed on the most widely employed techniques to probe the LSS, viz.\ the summary statistics introduced in \cref{sec:correlators}. In this section, instead, we present other promising probes, whilst keeping a clear SKAO perspective. 
For instance, the survey capabilities of the SKA-Mid, tracing redshifts of \hi\ galaxies and the variations in continuum brightness over cosmic time, enable a measurement of redshift drift \citep{kloeckner2015}, mapping the expansion history of the Universe in a model-independent way \citep{sandage1962}. 
The expected signal is a velocity shift of a few \(\mathrm{cm\,s^{-1}}\) over a timescale of approximately a decade. This requires precise determination of the line-centre offset of the averaged \hi\ emission line profile. A key factor is the number of \hi\ galaxies used in the stacking process. 

\begin{figure}
    \hspace{-1.8cm}
\includegraphics[width=1.2\textwidth]{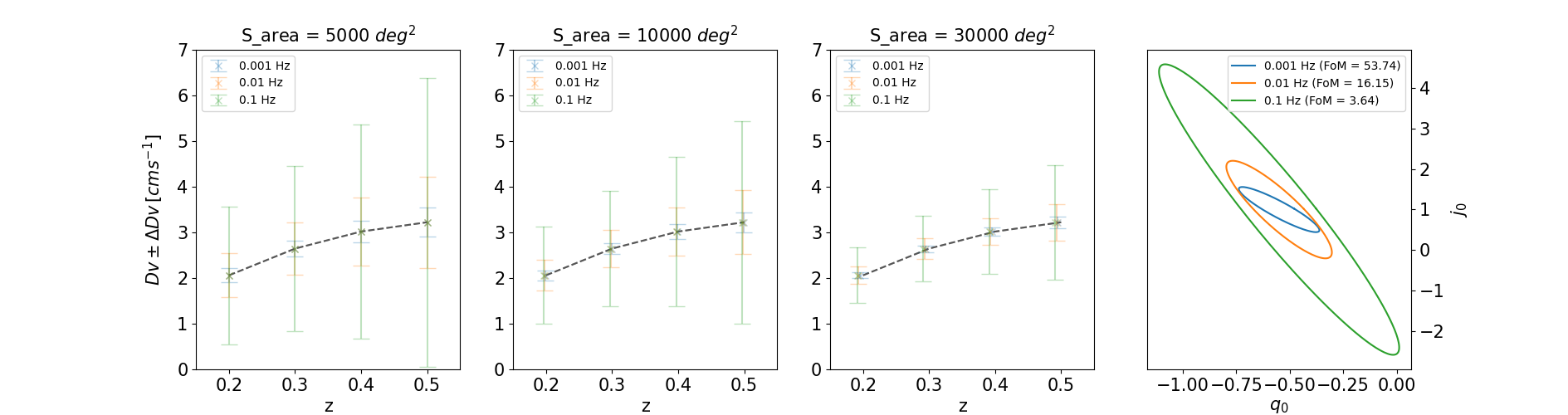}
\caption{\textit{Left plots:} Theoretical prediction and error estimates of the redshift drift versus redshift, for different channel width (\(0.001\), \(0.01\), and \(0.1\,\si{\hertz}\)) and survey areas (\(5\,000\), \(10\,000\), and \(30\,000\,\deg^2\)). \textit{Right-most plot:} Confidence regions for the cosmographic deceleration, and jerk parameters, assuming a prior in the marginalized Hubble constant, for the same choices of channel width and survey area of \(5\,000\,\deg^2\).}
    \label{fig:redshiftdrift1}
\end{figure}

To estimate uncertainties in the line-centre offset, we assume that individual \hi\ emission lines are Gaussian distributed with no significant technical systematics. The uncertainty in the redshift drift can then be described by the error in the line centre of a Gaussian profile \cite[][Eq.\ 14]{minin2009}, weighted by the square root of the number of stacked spectra.
\Cref{fig:redshiftdrift1} (three panels from the left) shows the expected redshift drift in units of $\mathrm{cm\,s^{-1}}$ within twelve years, with estimated SKA-Mid AA$^\ast$ errors for several choices of channel width and sky coverage, assuming one hour of observation per pointing. 
Even with a moderate sky coverage of \(5\,000\,\deg^2\) and a survey duration of approximately six months, a redshift-drift experiment is within reach of the SKA-Mid AA$^\ast$. Here, the average uncertainty in the \hi-line is fixed to \(\qty{150}{\kilo\meter\per\second}\) \citep[SAX simulations,][]{obreschkow2009}, whilst the angular number density of \hi\ galaxies uses the updated functional description of the number densities of \citet{yahya2015}. We further assume single-channel detections from previous large-sky surveys. The values of the MPA are defined by the flux-density boundaries given in \citet{yahya2015}. 

Constraints from these measurements can be forecast using information matrix analysis tools \citep{alves2019,Rocha2022,marques2023}, which directly constrain the cosmographic series up to the jerk parameter, i.e.\ $H_0$, $q_0$, and $j_0$ (see also \cref{sec:mg}). 
With a conservative prior $\sigma(H_0)=\qty{10}{\kilo\meter\per\second\per\mega\parsec}$, constraints in the $(q_0,j_0)$-plane are presented in the rightmost panel of \cref{fig:redshiftdrift1}, including a figure of merit (FoM), defined as the inverse of the area of the one-sigma confidence ellipse. 
We use the publicly available {\tt FRIDDA} code and assume three redshift bins centred at \(0.1\), \(0.3\), and \(0.5\), considering velocity uncertainties corresponding to the scenarios in \cref{fig:redshiftdrift1}. We find one-sigma uncertainties for $q_0$ and $j_0$ ranging from \(0.65\) to \(0.08\) and from \(4.2\) to \(0.25\) respectively. This shows that with a reasonable investment of observing time, SKA-Mid AA$^\ast$ can deliver meaningful model-independent constraints. The best-fit concordance model yields model-dependent predictions for them, to be compared to these observations, validating or excluding the model. Such early observations will also clarify technical requirements, identify systematic bottlenecks, and optimize subsequent observational strategies. Since the redshift drift signal grows linearly with time, the precision of such measurements will improve with extended observational baselines, making SKA-Mid a powerful tool for future cosmographic studies.
%

Furthermore, gravity theories can also be tested through gravitational wave observations. The detection of polarisation modes other than the two tensor modes may signal deviations from Einstein's theory of gravity \citep{Eardley_1973a,Eardley_1973b}. Moreover, these modes can be frequency-dependent \citep{Hyun_2019}, and modified gravity models can accommodate parity violation features such as amplitude and velocity birefringence effects in their dispersion relations. In fact, the ripples in the fabric of spacetime exist across a range of frequencies, or wavelengths. Low-frequency gravitational wave signals can form a stochastic background and originate either from the mergers of supermassive black hole binaries or from early time processes such as the ones which can arise from inflationary processes including the reheating and preheating periods where cosmological phase transitions can occur. These signals produce a distinctive signature, namely they show a quadrupolar spatial correlation between the arrival time of pulses emitted by different millisecond pulsar pairs, associated with the Hellings-Downs curve \citep{HellingsDowns_1983}. These millisecond pulsars are not likely to be sensitive to starquakes and glitches, thus acting as stable astrophysical clocks and reliable laboratories to test GWs \citep{Verbiest:2021}. Therefore, scientific missions using pulsar timing techniques have been implemented.
 
The SKAO will reach peaks of sensitivity higher than those of current pulsar timing array experiments, such as the International Pulsar Timing Array consortium \citep{IPTA_2019}. Moreover, it will enable probes of the stochastic gravitational wave background and searches for individual supermassive black hole binaries. This complements other ground- and space-based GW experiments, such as the ET \citep{ET_2025}, the LIGO-Virgo-Kagra Collaboration \citep{LVK_4OR2025}, and LISA \citep{LISA_2023}. This wide range of frequencies covered by these observatories will provide insights into the nature of gravity, as most of these experiments and the GW sources they probe do not overlap, hence the information extracted from each will be different and complementary.

\section{Future perspectives}\label{sec:SKA2}
%
A future upgrade of the SKAO, often referred to as `Phase 2' of the SKAO, will enable additional tests of fundamental physics on cosmological scales \citep[e.g.][]{2020PASA...37....2W}. Thanks to the increased statistics in the number of spectroscopically observed \hi-galaxies, it will provide sensitivity to GR effects in the clustering signal. As discussed in \cref{sec:rel_effects}, these effects generate a dipole term in the two-point correlation function of galaxies, which is expected to be detectable with a signal-to-noise of around 80 by Phase 2 of the SKAO \citep{Castello:2023zjr}.

One of these GR effects is gravitational redshift, which corresponds to the frequency change of photons crossing gravitational potential differences. As such, this effect is directly sensitive to the time distortion $\Psi$, hence giving access to one of the four fundamental fields in \cref{fig:four_fields} and providing crucial information for testing the properties of gravity and dark matter on cosmological scales. To achieve this, gravitational redshift can be combined with RSDs, which probe peculiar velocities $V$, and weak gravitational lensing, which is sensitive to the sum of the two potentials $\Phi+\Psi$. In the following, we illustrate the new tests that can be performed with these combinations of observables. 



RSDs have been used to constrain modifications to the Poisson equation parametrised through the phenomenological function $\mu$ in \cref{fig:four_fields} \citep[see e.g.][]{eBOSS:2020yzd}. Such constraints are obtained by employing the Euler equation to relate the peculiar velocities measured from the RSDs to the time distortion $\Psi$, which can then be combined with measurements of the density $\delta$ from galaxy clustering, as per \cref{fig:four_fields}. However, \citet{Castello:2022uuu} (see also \citealt{Bonvin:2023ghp, Castello:2024jmq}) showed that this approach is affected by a dangerous shortcoming, as it relies on the restrictive assumption that CDM obeys the weak equivalence principle, encoded in the Euler equation. Physically, this corresponds to the statement that CDM falls into gravitational potentials in the same way as ordinary matter, which has not been proven to date and does not hold in several extended dark matter models. This could be due to gravity coupling differently to different matter species \citep[e.g.][]{Gleyzes:2015pma} or arising as an effective violation due to non-gravitational interactions of CDM with Standard-Model particles \citep[e.g.][]{Barkana:2018lgd} or within the dark sector  \citep{Pourtsidou:2013nha, Spergel:1999mh, Tulin:2017ara}.

However, without the restrictive assumption of the weak equivalence principle, RSD analyses show a strong degeneracy between the function $\mu$ in the Poisson equation and modifications to the Euler equation \citep{Castello:2022uuu}. Such modifications can be parameterised with two additional time- and scale-dependent phenomenological functions: $\Theta$ and $\Gamma$ in \cref{fig:four_fields}, representing a fifth force and a friction term acting on CDM, respectively \citep{Bonvin:2018ckp}.
By probing the growth rate of cosmic structure, RSD surveys can only constrain the fully degenerate combination $\mu+\Gamma$ and thus cannot distinguish between the effect of $\mu$, encoding the depth of gravitational potentials, and that of $\Gamma$, controlling how CDM falls into these potentials. Such modifications can have the very same impact on the evolution of cosmic structure. Additionally, there is a strong degeneracy with the friction term $\Theta$, leading to a severe widening of the joint constraint on $\mu+\Gamma$.

Luckily, including measurements of gravitational redshift from SKAO Phase 2 efficiently lifts this degeneracy, leading to tight individual constraints on gravity modifications and violations of the weak equivalence principle. Notably, the value of $\mu$ at present time is forecasted to be constrained at the level of $14.7\%$, exceeding the precision of current RSD constraints obtained under the limiting assumption that the weak equivalence principle is valid \citep{eBOSS:2020yzd}. This remarkable result is motivated by the fact that gravitational redshift provides complementary information to RSDs, as it probes how photons escape from a gravitational potential, while RSDs are sensitive to the growth of matter perturbations.  This is crucial to disentangle modified gravity from extended CDM scenarios where the weak equivalence principle is broken, which would otherwise be completely indistinguishable in a standard RSD analysis \citep{Bonvin:2023ghp,Castello:2024jmq}.

\cite{Castello:2023zjr} further demonstrated the constraining power of gravitational redshift with a complementary approach, based on theoretically motivated parameters. This analysis was performed within the framework of the effective theory of interacting dark energy, a broad class of extended models covering the full set of Horndeski theories and at the same time allowing for different couplings of the cosmic components to gravity. Also in this case, the inclusion of gravitational redshift in the forecasts for the SKAO Phase 2 leads to a clear breaking of degeneracies in the parameter space and improvements in the constraints of up to 50\%. 

\begin{figure}
    \centering
\includegraphics[width=0.7\textwidth]{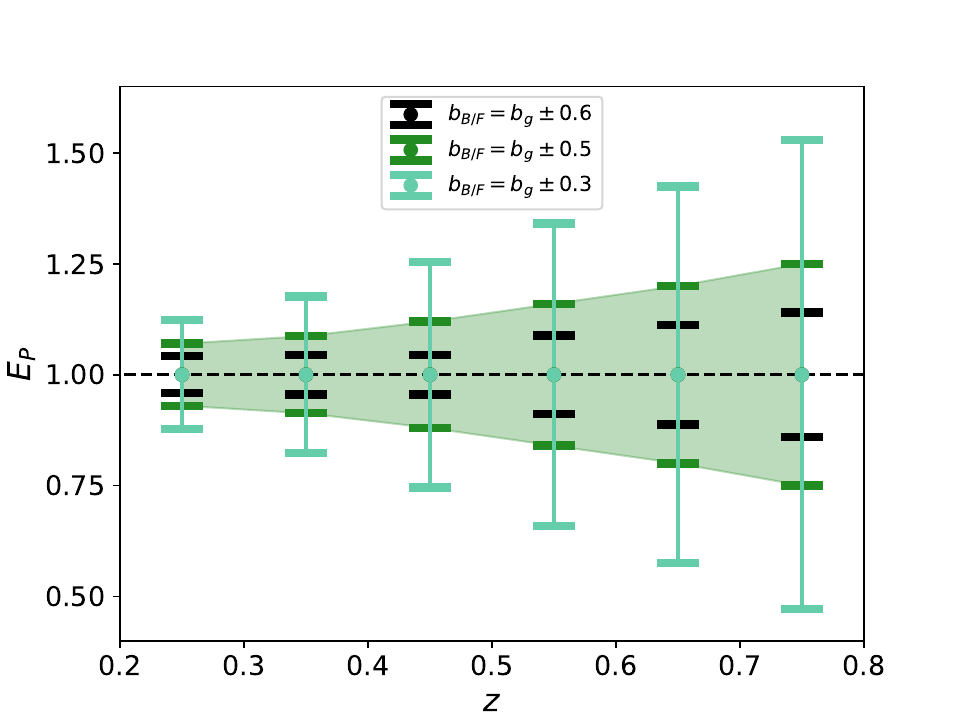}
    \caption{Constraints on deviations from the weak equivalence principle from SKAO Phase 2, encoded in the deviation of the parameter $E_P$ from 1 as a function of redshift. The colours correspond to various bias differences between the two galaxy populations (bright B and faint F) considered to detect the dipole in the 2-pt correlation function, which is sensitive to GR effects as discussed in Section \ref{sec:rel_effects}. Larger bias differences improve the detection significance and hence the constraints. (Reproduced from \citealt{Castello:2024lhl}.)
    }
    \label{fig:EP_constraints}
\end{figure}

The combination of RSDs and gravitational redshift also provides a direct test of the Euler equation by linking $\Psi$ to $V$, as can be seen from \cref{fig:four_fields}. \cite{Castello:2024lhl} developed a model-independent null test $E_P$ that deviates from 1 in the presence of a violation, which relies on minimal assumptions and in particular does not require specifying the power spectrum shape, the background evolution, the growth rate of cosmic structure, the galaxy bias or a model for the potential violation of the equivalence principle. Forecasts for SKAO Phase 2 indicated expected constraints around $10$--$15\%$ as a function of redshift, as shown in \cref{fig:EP_constraints}.
%

Although gravitational redshift with SKAO Phase 2 will give us access to the time distortion $\Psi$, weak lensing surveys provide highly complementary information, as they are sensitive to the sum $\Phi+\Psi$. As suggested in \cite{Sobral-Blanco:2021cks} and further explored in \cite{Tutusaus:2022cab}, the combination of gravitational redshift and lensing can thus be exploited to constrain anisotropic stress, defined as the ratio $\eta\equiv \Phi/\Psi$. This presents a direct comparison of the two metric degrees of freedom, which are predicted to be equal if general relativity holds. As such, it is a key test of the governing theory of gravity. \cite{Tutusaus:2022cab} has found that, with a combination of gravitational redshift from SKAO Phase 2 and gravitational lensing from LSST, the quantity $2/(1+\eta)$ can be constrained at the level of $\sim20\%$ in an optimistic scenario (and $\sim 30\%$ in the pessimistic case). This constraint is obtained in a model-independent manner, directly comparing the two metric components without relying on assumptions such as the validity of the weak equivalence principle. 


\section{Conclusions}\label{sec:conclusions}
The coming decade will see a step change in our ability to test gravity on cosmological scales, the SKAO playing a central role. By mapping \hi-line and radio-continuum galaxies over huge comoving volumes and a broad redshift range, the SKAO will probe both the background expansion and the growth of structure in a way that is highly complementary to optical/near-infrared observations and CMB surveys. Within the \lcdm\ framework, this will allow for stringent internal consistency tests: does a single relativistic model, with a small set of parameters, account simultaneously for baryon acoustic oscillations, RSDs, weak lensing, and higher-order clustering across radio, optical and microwave data sets? A positive answer would significantly strengthen the status of GR+\lcdm\ as an effective theory on ultra-large scales; a negative answer would help localise where the standard picture fails, whether in the expansion history, in the growth of perturbations, or in the relativistic projection effects that link theory to observables.

A key theme of this chapter has been the role of intrinsically relativistic effects in LSS observables (\cref{sec:rel_effects}). These have traditionally been treated as small `corrections' to Newtonian analyses, lying just beyond the reach of current data. SKAO surveys will change this situation. The combination of very large survey volumes, multi-tracer strategies (e.g.\ faint-bright splits, radio-optical cross-correlations), and measurements in both configuration and Fourier space opens up realistic prospects for detecting Doppler terms, gravitational redshift signatures, lensing magnification, and other light-cone effects. These signals are not merely nuisances to be modelled away: they are direct tests of GR in the ultra-weak field regime. At the same time, a robust treatment of these terms will be essential for unbiased constraints on PNG and on any scale-dependent deviations from the standard picture.

Beyond testing GR, the SKAO will provide a powerful laboratory for alternatives and extensions to the \lcdm\ model (\cref{sec:mg}). Model-independent frameworks such as the effective field theory of dark energy and phenomenological parameterisations of modified gravity (e.g.\ $\mu$, $\eta$, $\Sigma$, and the growth index $\gamma$) offer a common language in which SKAO measurements can be confronted with those from \textit{Planck}, \textit{Euclid}, DESI, LSST, and future CMB experiments. \hi\ intensity mapping and radio galaxy clustering will sharpen constraints on the time and scale dependence of the effective gravitational coupling and lensing potential, while higher-order statistics (bispectra and beyond) will help to disentangle relativistic projection effects from primordial signatures such as local $f_{\rm NL}$.

Looking ahead, Phase~2 of the SKAO (\cref{sec:SKA2}) will extend these capabilities substantially, turning several of the forecasts in this chapter into precision tests. A high-significance detection of the relativistic dipole in the 2- and 3-point correlation functions, direct constraints on gravitational redshift and anisotropic stress, and model-independent tests of the Euler equation and the weak equivalence principle become realistic targets. In combination with optical, infrared, CMB and gravitational-wave facilities, the SKAO will allow us to improve significantly on standard cosmological gravity tests: either reinforcing GR+\lcdm\ as an excellent effective description across a wide dynamical range or providing clear signposts towards new physics in the dark sector or in the law of gravity itself. In either outcome, radio cosmology with the SKAO will be central to our attempt to understand the origin of cosmic acceleration and the behaviour of gravity on the largest scales accessible to observation.

\section*{Author List Ordering}
Authors for this chapter are ordered as: corresponding author, alphabetical tiers. Authors in the first tier contributed to the chapter by producing original results using SKAO updated specifications. The second tier comprises authors who wrote significant part of the text and/or provided necessary inputs for the analysis. Authors in the last tier contributed with useful comments and constructive feedback, active supervision, and valuable inputs for discussion.

\section*{Acknowledgments}
StC, SJR, and FM acknowledge support from the Italian Ministry of University and Research (\textsc{mur}), PRIN 2022 `EXSKALIBUR – Euclid-Cross-SKA: Likelihood Inference Building for Universe's Research', Grant No.\ 20222BBYB9, CUP D53D2300252 0006, from the Italian Ministry of Foreign Affairs and International Cooperation (\textsc{maeci}), Grant No.\ ZA23GR03, and from the European Union -- Next Generation EU. TM is supported by funding from the European Research Council (ERC) under the European Union’s HORIZON-ERC-2022 (Grant No.\ 101076865). SvC, NG, and CB acknowledge funding from the European Research Council (ERC) under the European Union's Horizon 2020 research and innovation program (Grant No.\ 863929; project title ``Testing the law of gravity with novel large-scale structure observables''). SvC is supported by a Postdoc Mobility Fellowship of the Swiss National Science Foundation (SNSF), project No.\ P500PT 230281. NG is supported by the STFC, Grant No.\ ST/B001175/1. SLG and RM are supported by the South African Radio Astronomy Observatory and the National Research Foundation, Grant No.\ 75415. CH's work is funded by the Volkswagen Foundation. This work was supported by the DFG under Germany's Excellence Strategy EXC 2181/1 -- 390900948 The Heidelberg STRUCTURES Excellence Cluster. This work was financed by Portuguese funds through FCT (Funda\c c\~ao para a Ci\^encia e a Tecnologia) in the framework of the project 2022.04048.PTDC (Phi in the Sky, DOI 10.54499/2022.04048.PTDC). CJAPM also acknowledges FCT and POCH/FSE (EC) support through Investigador FCT Contract 2021.01214.CEECIND/CP1658/CT0001 (DOI 10.54499/2021.01214.CEECIND/CP1658/CT0001). ZS acknowledges support from the research projects PID2021-123012NB-C43, PID2024-159420NB-C43, the Proyecto de Investigaci\'on SAFE 25003 from the Consejo Superior de Investigaciones Cient\'ificas (CSIC), and the Spanish Research Agency (Agencia Estatal de Investigaci\'on) through the Grant IFT Centro de Excelencia Severo Ochoa No.\ CEX2020-001007-S, funded by MCIN/AEI/10.13039/501100011033.

\bibliographystyle{abbrvnat-maxbibnames4}
\bibliography{chapter} 

\end{document}

%% file: plots/Fig1.tex
\begin{figure}
    \centering

    \begin{tikzpicture}[
        font=\small,
        var/.style={align=center},
        eqn/.style={font=\small, fill=white, inner sep=1pt},
        arrow/.style={<->, very thick, >=Latex},
        orangearrow/.style={arrow, orange!90!black},
        greenarrow/.style={arrow, green!60!black},
        bluearrow/.style={arrow, blue!70!black},
    ]
    \centering
    
    \node[font=\bfseries\large] at (3.7,6.6) {Standard scenario};
    \node[font=\bfseries\large] at (10.5,6.6) {
    Modified scenario};
    
    \node[align=center] at (1.8,5.7) {Density};
    \node[align=center] at (5.5,5.7) {Velocity};
    \node[var] (deltaL) at (1.8,5) {\LARGE{$\delta$}};
    \node[var] (vL) at (5.5,5) {\LARGE{$V$}};
    \node[var] (phiL) at (1.8,1.8) {\LARGE{$\Phi$}};
    \node[var] (psiL) at (5.5,1.8) {\LARGE{$\Psi$}};
    
    \draw[bluearrow] (vL) -- (deltaL) 
        node[midway, eqn, align=center, draw, line width=0.2mm, inner sep=2pt] 
        {Continuity \\ [-5pt] equation};
    
    \draw[bluearrow] (vL) -- (psiL) 
        node[midway, eqn, align=center, draw, line width=0.2mm, inner sep=2pt] 
        {Euler \\ [-5pt] equation};
    
    \draw[bluearrow] (phiL) -- (deltaL) 
        node[midway, eqn, align=center, draw, line width=0.2mm, inner sep=2pt] 
        {Poisson \\ [-5pt] equation};
    
    \node at (3.65,1.8) {\LARGE \textcolor{blue}{$=$}};
    
    \node[align=center] at (1.8,0.9) {Space \\ [-3pt] distortion};
    \node[align=center] at (5.5,0.9) {Time \\ [-3pt] distortion};
    
    \node[var] (deltaR) at (8.5,5) {\LARGE{$\delta$}};
    \node[var] (vR) at (12.4,5) {\LARGE{$V$}};
    \node[var] (phiR) at (8.5,1.8) {\LARGE{$\Phi$}};
    \node[var] (psiR) at (12.4,1.8) {\LARGE{$\Psi$}};
    
    \draw[greenarrow] (vR) -- (deltaR) 
        node[midway, eqn, align=center, draw, line width=0.2mm, inner sep=2pt] 
        {\textcolor{green!60!black}{\parbox{
        0.6cm}{\centering
        \large{$K$}}}};
        
    \draw[greenarrow] (vR) -- (psiR) node[midway, eqn, align=center, draw, line width=0.2mm, inner sep=2pt]  {\textcolor{green!60!black}{\parbox{
    1cm}{\centering
    \large{$\Theta$, $\Gamma$}}}};
    
    \draw[orangearrow] (psiR) -- (deltaR) node[midway, eqn, align=center, draw, line width=0.2mm, inner sep=2pt] {\large \textcolor{orange!90!black}{\parbox{0.5cm}{ \centering \large{$\mu$}}}};
    
    \node at (10.35, 2.3) {\large \textcolor{orange!90!black}{\boxed{\eta}}};
    \node at (10.35,1.8) {\LARGE \textcolor{orange!90!black}{$\neq$}};
    
    \node[align=center, text=orange!90!black] at (10.35,1.1) {Modified gravity};
    \node[align=center, text=green!60!black] at (10.35,0.7) {Non-standard dark matter};
    
    \node[align=right, font=\itshape] at (0,5) {Matter 
    fields};
    \node[align=right, font=\itshape] at (0,1.8) {Gravitational 
    \\ [-2pt]
    potentials};
    \end{tikzpicture}
    
    \caption[Four fields]{Fundamental relations between the matter fields and the gravitational potentials in the standard cosmological model (left) and with generic deviations due to gravity modifications and non-standard dark matter properties (right). The deviations on the right panel are generic free functions of redshift and scale. (Reproduced from \citealt{Castello:2025qsm}.) 
    }
    \label{fig:four_fields}
\end{figure}

%% file: main.bbl
\begin{thebibliography}{147}
\providecommand{\natexlab}[1]{#1}
\providecommand{\url}[1]{\texttt{#1}}
\expandafter\ifx\csname urlstyle\endcsname\relax
  \providecommand{\doi}[1]{doi: #1}\else
  \providecommand{\doi}{doi: \begingroup \urlstyle{rm}\Url}\fi

\bibitem[Abac et~al.(2025{\natexlab{a}})]{ET_2025}
A.~Abac et~al.
\newblock 3 2025{\natexlab{a}}.

\bibitem[Abac et~al.(2025{\natexlab{b}})]{LVK_4OR2025}
A.~G. Abac et~al.
\newblock 8 2025{\natexlab{b}}.

\bibitem[Abramo and Leonard(2013)]{Abramo_2013}
L.~R. Abramo and K.~E. Leonard.
\newblock \emph{\mnras}, 432\penalty0 (1):\penalty0 318–326, Apr. 2013.
\newblock ISSN 1365-2966.
\newblock \doi{10.1093/mnras/stt465}.
\newblock URL \url{http://dx.doi.org/10.1093/mnras/stt465}.

\bibitem[Abramo et~al.(2022)Abramo, Ferri, and Tashiro]{Abramo_2022}
L.~R. Abramo, J.~V.~D. Ferri, and I.~L. Tashiro.
\newblock \emph{\jcap}, 2022\penalty0 (04):\penalty0 013, Apr. 2022.
\newblock ISSN 1475-7516.
\newblock \doi{10.1088/1475-7516/2022/04/013}.
\newblock URL \url{http://dx.doi.org/10.1088/1475-7516/2022/04/013}.

\bibitem[{Addis} et~al.(2025){Addis}, {Guandalin}, and {Clarkson}]{Addis_2025}
C.~{Addis}, C.~{Guandalin}, and C.~{Clarkson}.
\newblock \emph{\jcap}, 2025\penalty0 (4):\penalty0 080, Apr. 2025.
\newblock \doi{10.1088/1475-7516/2025/04/080}.

\bibitem[Addis et~al.(2025)Addis, Guedezounme, Hammond, Clarkson, Montano,
  Camera, Jolicoeur, and Maartens]{Addis_2025_new}
C.~Addis et al.
\newblock 2025.
\newblock URL \url{https://arxiv.org/abs/2511.09466}.

\bibitem[Alam et~al.(2021)]{eBOSS:2020yzd}
S.~Alam et~al.
\newblock \emph{Phys. Rev. D}, 103\penalty0 (8):\penalty0 083533, 2021.
\newblock \doi{10.1103/PhysRevD.103.083533}.

\bibitem[Alonso and Ferreira(2015)]{Alonso:2015sfa}
D.~Alonso and P.~G. Ferreira.
\newblock \emph{Phys. Rev. D}, 92\penalty0 (6):\penalty0 063525, 2015.
\newblock \doi{10.1103/PhysRevD.92.063525}.

\bibitem[Alves et~al.(2019)Alves, Leite, Martins, Matos, and Silva]{alves2019}
C.~S. Alves et al.
\newblock \emph{\mnras}, 488\penalty0 (3):\penalty0 3607--3624, 07 2019.

\bibitem[Amendola et~al.(2008)Amendola, Kunz, and Sapone]{Amendola:2007rr}
L.~Amendola, M.~Kunz, and D.~Sapone.
\newblock \emph{JCAP}, 04:\penalty0 013, 2008.
\newblock \doi{10.1088/1475-7516/2008/04/013}.

\bibitem[Asorey\&Hale et~al.(2026)Asorey\&Hale, author2, author3, author4, and
  author5]{Asorey01.2026.SKA}
J.~Asorey\&Hale et al.
\newblock In \emph{Advancing Astrophysics with the SKA -- II (AASKAII)}. 2026.
\newblock arXiv search: Report number AASKAII/Asorey01.

\bibitem[Assassi et~al.(2017)Assassi, Simonovi{\'c}, and
  Zaldarriaga]{Assassi:2017lea}
V.~Assassi, M.~Simonovi{\'c}, and M.~Zaldarriaga.
\newblock \emph{JCAP}, 11:\penalty0 054, 2017.
\newblock \doi{10.1088/1475-7516/2017/11/054}.

\bibitem[Auclair et~al.(2023)]{LISA_2023}
P.~Auclair et~al.
\newblock \emph{Living Rev. Rel.}, 26\penalty0 (1):\penalty0 5, 2023.
\newblock \doi{10.1007/s41114-023-00045-2}.

\bibitem[Babichev and Deffayet(2013)]{Babichev:2013usa}
E.~Babichev and C.~Deffayet.
\newblock \emph{Class. Quant. Grav.}, 30:\penalty0 184001, 2013.
\newblock \doi{10.1088/0264-9381/30/18/184001}.

\bibitem[{Bacon} et~al.(2014){Bacon}, {Andrianomena}, {Clarkson}, {Bolejko},
  and {Maartens}]{2014MNRAS.443.1900B}
D.~J. {Bacon} et al.
\newblock \emph{\mnras}, 443\penalty0 (3):\penalty0 1900--1915, Sept. 2014.
\newblock \doi{10.1093/mnras/stu1270}.

\bibitem[Baker et~al.(2021)]{Baker:2019gxo}
T.~Baker et~al.
\newblock \emph{Rev. Mod. Phys.}, 93\penalty0 (1):\penalty0 015003, 2021.
\newblock \doi{10.1103/RevModPhys.93.015003}.

\bibitem[Barkana(2018)]{Barkana:2018lgd}
R.~Barkana.
\newblock \emph{Nature}, 555\penalty0 (7694):\penalty0 71--74, 2018.
\newblock \doi{10.1038/nature25791}.

\bibitem[Beltr{\'a}n~Jim{\'e}nez et~al.(2019)Beltr{\'a}n~Jim{\'e}nez,
  Heisenberg, and Koivisto]{BeltranJimenez:2019esp}
J.~Beltr{\'a}n~Jim{\'e}nez, L.~Heisenberg, and T.~S. Koivisto.
\newblock \emph{Universe}, 5\penalty0 (7):\penalty0 173, 2019.
\newblock \doi{10.3390/universe5070173}.

\bibitem[{Bernardeau} et~al.(2010){Bernardeau}, {Bonvin}, and
  {Vernizzi}]{2010PhRvD..81h3002B}
F.~{Bernardeau}, C.~{Bonvin}, and F.~{Vernizzi}.
\newblock \emph{\prd}, 81\penalty0 (8):\penalty0 083002, Apr. 2010.
\newblock \doi{10.1103/PhysRevD.81.083002}.

\bibitem[{Bernstein} and {Jain}(2004)]{2004ApJ...600...17B}
G.~{Bernstein} and B.~{Jain}.
\newblock \emph{\apj}, 600\penalty0 (1):\penalty0 17--25, Jan. 2004.
\newblock \doi{10.1086/379768}.

\bibitem[Berti et~al.(2022)Berti, Spinelli, Haridasu, Viel, and
  Silvestri]{Berti:2021ccw}
M.~Berti et al.
\newblock \emph{JCAP}, 01\penalty0 (01):\penalty0 018, 2022.
\newblock \doi{10.1088/1475-7516/2022/01/018}.

\bibitem[{Beutler} and {Di Dio}(2020)]{Beutler_2020}
F.~{Beutler} and E.~{Di Dio}.
\newblock \emph{\jcap}, 2020\penalty0 (7):\penalty0 048, July 2020.
\newblock \doi{10.1088/1475-7516/2020/07/048}.

\bibitem[{Bharadwaj} et~al.(2001){Bharadwaj}, {Nath}, and
  {Sethi}]{Bharadwaj2001}
S.~{Bharadwaj}, B.~B. {Nath}, and S.~K. {Sethi}.
\newblock \emph{Journal of Astrophysics and Astronomy}, 22\penalty0
  (1):\penalty0 21--34, Mar. 2001.
\newblock \doi{10.1007/BF02933588}.

\bibitem[Bonvin and Durrer(2011)]{Bonvin:2011bg}
C.~Bonvin and R.~Durrer.
\newblock \emph{Phys. Rev. D}, 84:\penalty0 063505, 2011.
\newblock \doi{10.1103/PhysRevD.84.063505}.

\bibitem[Bonvin and Fleury(2018)]{Bonvin:2018ckp}
C.~Bonvin and P.~Fleury.
\newblock \emph{JCAP}, 05:\penalty0 061, 2018.
\newblock \doi{10.1088/1475-7516/2018/05/061}.

\bibitem[Bonvin and Pogosian(2023)]{Bonvin:2023ghp}
C.~Bonvin and L.~E. Pogosian.
\newblock \emph{Nature Astron.}, 7\penalty0 (9):\penalty0 1023--1024, 2023.
\newblock \doi{10.1038/s41550-023-02026-5}.

\bibitem[Bonvin et~al.(2014)Bonvin, Hui, and Gaztanaga]{Bonvin:2013ogt}
C.~Bonvin, L.~Hui, and E.~Gaztanaga.
\newblock \emph{Phys. Rev. D}, 89\penalty0 (8):\penalty0 083535, 2014.
\newblock \doi{10.1103/PhysRevD.89.083535}.

\bibitem[Bruni et~al.(2012)Bruni, Crittenden, Koyama, Maartens, Pitrou, and
  Wands]{Bruni:2011ta}
M.~Bruni et al.
\newblock \emph{Phys. Rev. D}, 85:\penalty0 041301, 2012.
\newblock \doi{10.1103/PhysRevD.85.041301}.

\bibitem[{Camera} et~al.(2015){Camera}, {Maartens}, and
  {Santos}]{2015MNRAS.451L..80C}
S.~{Camera}, R.~{Maartens}, and M.~G. {Santos}.
\newblock \emph{\mnras}, 451:\penalty0 L80--L84, July 2015.
\newblock \doi{10.1093/mnrasl/slv069}.

\bibitem[Camera et~al.(2015)]{Camera:2015yqa}
S.~Camera et~al.
\newblock \emph{PoS}, AASKA14:\penalty0 025, 2015.
\newblock \doi{10.22323/1.215.0025}.

\bibitem[Casas et~al.(2023)Casas, Carucci, Pettorino, Camera, and
  Martinelli]{Casas:2022vik}
S.~Casas et al.
\newblock \emph{Phys. Dark Univ.}, 39:\penalty0 101151, 2023.
\newblock \doi{10.1016/j.dark.2022.101151}.

\bibitem[Castello(2025)]{Castello:2025qsm}
S.~Castello.
\newblock \emph{{A Matter of Time: Testing Gravity and Dark Matter with the
  Distortion of Time in Galaxy Surveys}}.
\newblock PhD thesis, University of Geneva, 2025.

\bibitem[Castello et~al.(2022)Castello, Grimm, and Bonvin]{Castello:2022uuu}
S.~Castello, N.~Grimm, and C.~Bonvin.
\newblock \emph{Phys. Rev. D}, 106\penalty0 (8):\penalty0 083511, 2022.
\newblock \doi{10.1103/PhysRevD.106.083511}.

\bibitem[Castello et~al.(2024{\natexlab{a}})Castello, Mancarella, Grimm,
  Sobral-Blanco, Tutusaus, and Bonvin]{Castello:2023zjr}
S.~Castello et al.
\newblock \emph{JCAP}, 05:\penalty0 003, 2024{\natexlab{a}}.
\newblock \doi{10.1088/1475-7516/2024/05/003}.

\bibitem[Castello et~al.(2024{\natexlab{b}})Castello, Wang, Dam, Bonvin, and
  Pogosian]{Castello:2024jmq}
S.~Castello et al.
\newblock \emph{Phys. Rev. D}, 110\penalty0 (10):\penalty0 103523,
  2024{\natexlab{b}}.
\newblock \doi{10.1103/PhysRevD.110.103523}.

\bibitem[Castello et~al.(2025)Castello, Zheng, Bonvin, and
  Amendola]{Castello:2024lhl}
S.~Castello, Z.~Zheng, C.~Bonvin, and L.~Amendola.
\newblock \emph{Phys. Rev. D}, 111\penalty0 (12):\penalty0 123559, 2025.
\newblock \doi{10.1103/1my7-zklj}.

\bibitem[{Castorina} and {Di Dio}(2022)]{2022JCAP...01..061C}
E.~{Castorina} and E.~{Di Dio}.
\newblock \emph{\jcap}, 2022\penalty0 (1):\penalty0 061, Jan. 2022.
\newblock \doi{10.1088/1475-7516/2022/01/061}.

\bibitem[Castorina and White(2018)]{Castorina_2018}
E.~Castorina and M.~White.
\newblock \emph{\mnras}, Feb. 2018.
\newblock ISSN 1365-2966.
\newblock \doi{10.1093/mnras/sty410}.
\newblock URL \url{http://dx.doi.org/10.1093/mnras/sty410}.

\bibitem[{Castro} et~al.(2005){Castro}, {Heavens}, and
  {Kitching}]{2005PhRvD..72b3516C}
P.~G. {Castro}, A.~F. {Heavens}, and T.~D. {Kitching}.
\newblock \emph{\prd}, 72\penalty0 (2):\penalty0 023516, July 2005.
\newblock \doi{10.1103/PhysRevD.72.023516}.

\bibitem[Challinor and Lewis(2011)]{Challinor:2011bk}
A.~Challinor and A.~Lewis.
\newblock \emph{Phys. Rev. D}, 84:\penalty0 043516, 2011.
\newblock \doi{10.1103/PhysRevD.84.043516}.

\bibitem[Clarkson et~al.(2019)Clarkson, de~Weerd, Jolicoeur, Maartens, and
  Umeh]{Clarkson:2018dwn}
C.~Clarkson et al.
\newblock \emph{Mon. Not. Roy. Astron. Soc.}, 486\penalty0 (1):\penalty0
  L101--L104, 2019.
\newblock \doi{10.1093/mnrasl/slz066}.

\bibitem[{Clifton} et~al.(2012){Clifton}, {Ferreira}, {Padilla}, and
  {Skordis}]{2012PhR...513....1C}
T.~{Clifton}, P.~G. {Ferreira}, A.~{Padilla}, and C.~{Skordis}.
\newblock \emph{\physrep}, 513\penalty0 (1):\penalty0 1--189, Mar. 2012.
\newblock \doi{10.1016/j.physrep.2012.01.001}.

\bibitem[{Cunnington} et~al.(2020){Cunnington}, {Camera}, and
  {Pourtsidou}]{2020MNRAS.499.4054C}
S.~{Cunnington}, S.~{Camera}, and A.~{Pourtsidou}.
\newblock \emph{\mnras}, 499\penalty0 (3):\penalty0 4054--4067, Dec. 2020.
\newblock \doi{10.1093/mnras/staa2986}.

\bibitem[{Cunnington} et~al.(2023){Cunnington}, {Wolz}, {Bull}, {Carucci},
  {Grainge}, {Irfan}, {Li}, {Pourtsidou}, {Santos}, {Spinelli}, and
  {Wang}]{2023MNRAS.523.2453C}
S.~{Cunnington} et al.
\newblock \emph{\mnras}, 523\penalty0 (2):\penalty0 2453--2477, Aug. 2023.
\newblock \doi{10.1093/mnras/stad1567}.

\bibitem[Deffayet et~al.(2011)Deffayet, Gao, Steer, and
  Zahariade]{Deffayet:2011gz}
C.~Deffayet, X.~Gao, D.~A. Steer, and G.~Zahariade.
\newblock \emph{Phys. Rev. D}, 84:\penalty0 064039, 2011.
\newblock \doi{10.1103/PhysRevD.84.064039}.

\bibitem[{DES Collaboration: Abbott} et~al.(2025)]{2025arXiv250313632D}
{DES Collaboration: Abbott} et~al.
\newblock \emph{arXiv e-prints}, art. arXiv:2503.13632, Mar. 2025.
\newblock \doi{10.48550/arXiv.2503.13632}.

\bibitem[{DESI Collaboration: Abareshi} et~al.(2022)]{2022AJ....164..207D}
{DESI Collaboration: Abareshi} et~al.
\newblock \emph{\aj}, 164\penalty0 (5):\penalty0 207, Nov. 2022.
\newblock \doi{10.3847/1538-3881/ac882b}.

\bibitem[{DESI Collaboration: Abdul-Karim} et~al.(2025)]{2025arXiv250314745D}
{DESI Collaboration: Abdul-Karim} et~al.
\newblock \emph{arXiv e-prints}, art. arXiv:2503.14745, Mar. 2025.
\newblock \doi{10.48550/arXiv.2503.14745}.

\bibitem[{DESI Collaboration: Adame} et~al.(2025)]{2025JCAP...09..008A}
{DESI Collaboration: Adame} et~al.
\newblock \emph{\jcap}, 9:\penalty0 008, Sept. 2025.
\newblock \doi{10.1088/1475-7516/2025/09/008}.

\bibitem[Di~Dio et~al.(2014)Di~Dio, Durrer, Marozzi, and
  Montanari]{DiDio:2014lka}
E.~Di~Dio, R.~Durrer, G.~Marozzi, and F.~Montanari.
\newblock \emph{JCAP}, 12:\penalty0 017, 2014.
\newblock \doi{10.1088/1475-7516/2014/12/017}.
\newblock [Erratum: JCAP 06, E01 (2015)].

\bibitem[Di~Dio et~al.(2016)Di~Dio, Durrer, Marozzi, and
  Montanari]{DiDio:2015bua}
E.~Di~Dio, R.~Durrer, G.~Marozzi, and F.~Montanari.
\newblock \emph{JCAP}, 01:\penalty0 016, 2016.
\newblock \doi{10.1088/1475-7516/2016/01/016}.

\bibitem[Di~Dio et~al.(2017)Di~Dio, Perrier, Durrer, Marozzi,
  Moradinezhad~Dizgah, Nore\~na, and Riotto]{DiDio:2016gpd}
E.~Di~Dio et al.
\newblock \emph{JCAP}, 03:\penalty0 006, 2017.
\newblock \doi{10.1088/1475-7516/2017/03/006}.

\bibitem[Di~Dio et~al.(2019)Di~Dio, Durrer, Maartens, Montanari, and
  Umeh]{DiDio:2018unb}
E.~Di~Dio et al.
\newblock \emph{JCAP}, 04:\penalty0 053, 2019.
\newblock \doi{10.1088/1475-7516/2019/04/053}.

\bibitem[{Dlamini} et~al.(2024){Dlamini}, {Jolicoeur}, and
  {Maartens}]{2024EPJC...84...95D}
S.~{Dlamini}, S.~{Jolicoeur}, and R.~{Maartens}.
\newblock \emph{European Physical Journal C}, 84\penalty0 (1):\penalty0 95,
  Jan. 2024.
\newblock \doi{10.1140/epjc/s10052-024-12467-5}.

\bibitem[{Dvornik} et~al.(2023){Dvornik}, {Heymans}, {Asgari}, {Mahony},
  {Joachimi}, et~al.]{2023A&A...675A.189D}
A.~{Dvornik} et al.
\newblock \emph{\aap}, 675:\penalty0 A189, July 2023.
\newblock \doi{10.1051/0004-6361/202245158}.

\bibitem[Eardley et~al.(1973{\natexlab{a}})Eardley, Lee, and
  Lightman]{Eardley_1973a}
D.~M. Eardley, D.~L. Lee, and A.~P. Lightman.
\newblock \emph{Phys. Rev. D}, 8:\penalty0 3308--3321, Nov 1973{\natexlab{a}}.
\newblock \doi{10.1103/PhysRevD.8.3308}.
\newblock URL \url{https://link.aps.org/doi/10.1103/PhysRevD.8.3308}.

\bibitem[Eardley et~al.(1973{\natexlab{b}})Eardley, Lee, Lightman, Wagoner, and
  Will]{Eardley_1973b}
D.~M. Eardley et al.
\newblock \emph{Phys. Rev. Lett.}, 30:\penalty0 884--886, Apr
  1973{\natexlab{b}}.
\newblock \doi{10.1103/PhysRevLett.30.884}.
\newblock URL \url{https://link.aps.org/doi/10.1103/PhysRevLett.30.884}.

\bibitem[{Euclid Collaboration: Mellier} et~al.(2025)]{2025A&A...697A...1E}
{Euclid Collaboration: Mellier} et~al.
\newblock \emph{\aap}, 697:\penalty0 A1, May 2025.
\newblock \doi{10.1051/0004-6361/202450810}.

\bibitem[{Fonseca} and {Clarkson}(2021)]{2021JCAP...12..003F}
J.~{Fonseca} and C.~{Clarkson}.
\newblock \emph{\jcap}, 2021\penalty0 (12):\penalty0 003, Dec. 2021.
\newblock \doi{10.1088/1475-7516/2021/12/003}.

\bibitem[Fonseca et~al.(2015)Fonseca, Camera, Santos, and
  Maartens]{Fonseca:2015laa}
J.~Fonseca, S.~Camera, M.~Santos, and R.~Maartens.
\newblock \emph{Astrophys. J. Lett.}, 812\penalty0 (2):\penalty0 L22, 2015.
\newblock \doi{10.1088/2041-8205/812/2/L22}.

\bibitem[Fonseca et~al.(2026)Fonseca, author2, author3, author4, and
  author5]{Fonseca01.2026.SKA}
J.~Fonseca et al.
\newblock In \emph{Advancing Astrophysics with the SKA -- II (AASKAII)}. 2026.
\newblock arXiv search: Report number AASKAII/Fonseca01.

\bibitem[Frusciante and Perenon(2020)]{Frusciante:2019xia}
N.~Frusciante and L.~Perenon.
\newblock \emph{Phys. Rept.}, 857:\penalty0 1--63, 2020.
\newblock \doi{10.1016/j.physrep.2020.02.004}.

\bibitem[{Ghosh} et~al.(2018){Ghosh}, {Durrer}, and
  {Sellentin}]{2018JCAP...06..008G}
B.~{Ghosh}, R.~{Durrer}, and E.~{Sellentin}.
\newblock \emph{\jcap}, 2018\penalty0 (6):\penalty0 008, June 2018.
\newblock \doi{10.1088/1475-7516/2018/06/008}.

\bibitem[Gleyzes et~al.(2015)Gleyzes, Langlois, Mancarella, and
  Vernizzi]{Gleyzes:2015pma}
J.~Gleyzes, D.~Langlois, M.~Mancarella, and F.~Vernizzi.
\newblock \emph{JCAP}, 08:\penalty0 054, 2015.
\newblock \doi{10.1088/1475-7516/2015/08/054}.

\bibitem[{Grasshorn Gebhardt} and {Dor{\'e}}(2021)]{2021PhRvD.104l3548G}
H.~S. {Grasshorn Gebhardt} and O.~{Dor{\'e}}.
\newblock \emph{\prd}, 104\penalty0 (12):\penalty0 123548, Dec. 2021.
\newblock \doi{10.1103/PhysRevD.104.123548}.

\bibitem[{Hall} et~al.(2013){Hall}, {Bonvin}, and
  {Challinor}]{2013PhRvD..87f4026H}
A.~{Hall}, C.~{Bonvin}, and A.~{Challinor}.
\newblock \emph{\prd}, 87\penalty0 (6):\penalty0 064026, Mar. 2013.
\newblock \doi{10.1103/PhysRevD.87.064026}.

\bibitem[Harrison et~al.(2026{\natexlab{a}})Harrison, author2, author3,
  author4, and author5]{Harrison01.2026.SKA}
I.~Harrison et al.
\newblock In \emph{Advancing Astrophysics with the SKA -- II (AASKAII)}.
  2026{\natexlab{a}}.
\newblock arXiv search: Report number AASKAII/Harrison02.

\bibitem[Harrison et~al.(2026{\natexlab{b}})Harrison, author2, author3,
  author4, and author5]{Harrison02.2026.SKA}
I.~Harrison et al.
\newblock In \emph{Advancing Astrophysics with the SKA -- II (AASKAII)}.
  2026{\natexlab{b}}.
\newblock arXiv search: Report number AASKAII/Harrison01.

\bibitem[Hassan and Rosen(2012)]{Hassan:2011hr}
S.~F. Hassan and R.~A. Rosen.
\newblock \emph{Phys. Rev. Lett.}, 108:\penalty0 041101, 2012.
\newblock \doi{10.1103/PhysRevLett.108.041101}.

\bibitem[Heisenberg(2014)]{Heisenberg:2014rta}
L.~Heisenberg.
\newblock \emph{JCAP}, 05:\penalty0 015, 2014.
\newblock \doi{10.1088/1475-7516/2014/05/015}.

\bibitem[{Hellings} and {Downs}(1983)]{HellingsDowns_1983}
R.~W. {Hellings} and G.~S. {Downs}.
\newblock \emph{\apjl}, 265:\penalty0 L39--L42, Feb. 1983.
\newblock \doi{10.1086/183954}.

\bibitem[{Heneka} and {Amendola}(2018)]{2018JCAP...10..004H}
C.~{Heneka} and L.~{Amendola}.
\newblock \emph{\jcap}, 2018\penalty0 (10):\penalty0 004, Oct. 2018.
\newblock \doi{10.1088/1475-7516/2018/10/004}.

\bibitem[Horndeski(1974)]{Horndeski:1974wa}
G.~W. Horndeski.
\newblock \emph{Int. J. Theor. Phys.}, 10:\penalty0 363--384, 1974.
\newblock \doi{10.1007/BF01807638}.

\bibitem[{Hu} and {Jain}(2004)]{2004PhRvD..70d3009H}
W.~{Hu} and B.~{Jain}.
\newblock \emph{\prd}, 70\penalty0 (4):\penalty0 043009, Aug. 2004.
\newblock \doi{10.1103/PhysRevD.70.043009}.

\bibitem[Hyun et~al.(2019)Hyun, Kim, and Lee]{Hyun_2019}
Y.-H. Hyun, Y.~Kim, and S.~Lee.
\newblock \emph{Phys. Rev. D}, 99:\penalty0 124002, Jun 2019.
\newblock \doi{10.1103/PhysRevD.99.124002}.
\newblock URL \url{https://link.aps.org/doi/10.1103/PhysRevD.99.124002}.

\bibitem[{Ishak}(2019)]{2019LRR....22....1I}
M.~{Ishak}.
\newblock \emph{Living Reviews in Relativity}, 22\penalty0 (1):\penalty0 1,
  Dec. 2019.
\newblock \doi{10.1007/s41114-018-0017-4}.

\bibitem[{Ivezi{\'c}} et~al.(2019){Ivezi{\'c}}, {Kahn}, {Tyson}, {Abel},
  {Acosta}, et~al.]{2019ApJ...873..111I}
{\v{Z}}.~{Ivezi{\'c}} et al.
\newblock \emph{\apj}, 873\penalty0 (2):\penalty0 111, Mar. 2019.
\newblock \doi{10.3847/1538-4357/ab042c}.

\bibitem[Jacobson(2007)]{Jacobson:2007veq}
T.~Jacobson.
\newblock \emph{PoS}, QG-PH:\penalty0 020, 2007.
\newblock \doi{10.22323/1.043.0020}.

\bibitem[Jelic-Cizmek et~al.(2021)Jelic-Cizmek, Lepori, Bonvin, and
  Durrer]{Jelic-Cizmek:2020pkh}
G.~Jelic-Cizmek, F.~Lepori, C.~Bonvin, and R.~Durrer.
\newblock \emph{JCAP}, 04:\penalty0 055, 2021.
\newblock \doi{10.1088/1475-7516/2021/04/055}.

\bibitem[Jeong and Schmidt(2020)]{Jeong:2019igb}
D.~Jeong and F.~Schmidt.
\newblock \emph{Phys. Rev. D}, 102\penalty0 (2):\penalty0 023530, 2020.
\newblock \doi{10.1103/PhysRevD.102.023530}.

\bibitem[{Jeong} et~al.(2012){Jeong}, {Schmidt}, and
  {Hirata}]{2012PhRvD..85b3504J}
D.~{Jeong}, F.~{Schmidt}, and C.~M. {Hirata}.
\newblock \emph{\prd}, 85\penalty0 (2):\penalty0 023504, Jan. 2012.
\newblock \doi{10.1103/PhysRevD.85.023504}.

\bibitem[{Jolicoeur} et~al.(2021){Jolicoeur}, {Maartens}, {De Weerd}, {Umeh},
  {Clarkson}, and {Camera}]{2021JCAP...06..039J}
S.~{Jolicoeur} et al.
\newblock \emph{\jcap}, 2021\penalty0 (6):\penalty0 039, June 2021.
\newblock \doi{10.1088/1475-7516/2021/06/039}.

\bibitem[Kaiser(1987)]{Kaiser:1987qv}
N.~Kaiser.
\newblock \emph{Mon. Not. Roy. Astron. Soc.}, 227:\penalty0 1--27, 1987.
\newblock \doi{10.1093/mnras/227.1.1}.

\bibitem[{Karagiannis} et~al.(2022){Karagiannis}, {Maartens}, and
  {Randrianjanahary}]{2022JCAP...11..003K}
D.~{Karagiannis}, R.~{Maartens}, and L.~F. {Randrianjanahary}.
\newblock \emph{\jcap}, 2022\penalty0 (11):\penalty0 003, Nov. 2022.
\newblock \doi{10.1088/1475-7516/2022/11/003}.

\bibitem[Khoury and Weltman(2004)]{Khoury:2003aq}
J.~Khoury and A.~Weltman.
\newblock \emph{Phys. Rev. Lett.}, 93:\penalty0 171104, 2004.
\newblock \doi{10.1103/PhysRevLett.93.171104}.

\bibitem[{Kl\"ockner} et~al.(2015){Kl\"ockner}, {Obreschkow}, Martins,
  Raccanelli, Champion, Roy, Lobanov, Wagner, and Keller]{kloeckner2015}
H.-R. {Kl\"ockner} et al.
\newblock \emph{PoS}, AASKA14:\penalty0 027, 2015.
\newblock \doi{10.22323/1.215.0027}.

\bibitem[Langlois and Noui(2016)]{Langlois:2015cwa}
D.~Langlois and K.~Noui.
\newblock \emph{JCAP}, 02:\penalty0 034, 2016.
\newblock \doi{10.1088/1475-7516/2016/02/034}.

\bibitem[Linder(2005)]{Linder:2005in}
E.~V. Linder.
\newblock \emph{Phys. Rev. D}, 72:\penalty0 043529, 2005.
\newblock \doi{10.1103/PhysRevD.72.043529}.

\bibitem[{Liu} et~al.(2020){Liu}, {Heneka}, and
  {Amendola}]{2020JCAP...05..038L}
X.-W. {Liu}, C.~{Heneka}, and L.~{Amendola}.
\newblock \emph{\jcap}, 2020\penalty0 (5):\penalty0 038, May 2020.
\newblock \doi{10.1088/1475-7516/2020/05/038}.

\bibitem[{Loeb} and {Wyithe}(2008)]{Loeb2008}
A.~{Loeb} and J.~S.~B. {Wyithe}.
\newblock \emph{\prl}, 100\penalty0 (16):\penalty0 161301, Apr. 2008.
\newblock \doi{10.1103/PhysRevLett.100.161301}.

\bibitem[Maartens et~al.(2015)Maartens, Abdalla, Jarvis, and
  Santos]{Maartens:2015mra}
R.~Maartens, F.~B. Abdalla, M.~Jarvis, and M.~G. Santos.
\newblock \emph{PoS}, AASKA14:\penalty0 016, 2015.
\newblock \doi{10.22323/1.215.0016}.

\bibitem[Maartens et~al.(2021)Maartens, Jolicoeur, Umeh, De~Weerd, and
  Clarkson]{Maartens:2020jzf}
R.~Maartens et al.
\newblock \emph{JCAP}, 04:\penalty0 013, 2021.
\newblock \doi{10.1088/1475-7516/2021/04/013}.

\bibitem[Marques et~al.(2023)Marques, Martins, and Alves]{marques2023}
C.~M.~J. Marques, C.~J.~A.~P. Martins, and C.~S. Alves.
\newblock {FRIDDA: Fisher foRecast code for combIned reDshift Drift and Alpha}.
\newblock Astrophysics Source Code Library, record ascl:2305.001, May 2023.

\bibitem[McDonald(2009)]{McDonald_2009}
P.~McDonald.
\newblock \emph{Journal of Cosmology and Astroparticle Physics}, 2009\penalty0
  (11):\penalty0 026–026, Nov. 2009.
\newblock ISSN 1475-7516.
\newblock \doi{10.1088/1475-7516/2009/11/026}.
\newblock URL \url{http://dx.doi.org/10.1088/1475-7516/2009/11/026}.

\bibitem[{McDonald} and {Seljak}(2009)]{McDonald&Seljak2009}
P.~{McDonald} and U.~{Seljak}.
\newblock \emph{\jcap}, 2009\penalty0 (10):\penalty0 007, Oct. 2009.
\newblock \doi{10.1088/1475-7516/2009/10/007}.

\bibitem[Minin and Kamalabadi(2009)]{minin2009}
S.~Minin and F.~Kamalabadi.
\newblock \emph{\ao}, 48\penalty0 (36):\penalty0 6913, Dec. 2009.
\newblock \doi{10.1364/AO.48.006913}.

\bibitem[Montandon et~al.(2023)Montandon, Adamek, Hahn, Nore{\~n}a, Rampf,
  Stahl, and van Tent]{Montandon:2022ulz}
T.~Montandon et al.
\newblock \emph{JCAP}, 08:\penalty0 043, 2023.
\newblock \doi{10.1088/1475-7516/2023/08/043}.

\bibitem[Montandon et~al.(2025)Montandon, Di~Dio, Rampf, and
  Adamek]{Montandon:2025uul}
T.~Montandon, E.~Di~Dio, C.~Rampf, and J.~Adamek.
\newblock 1 2025.

\bibitem[{Montano} and {Camera}(2024{\natexlab{a}})]{Montano2024_1}
F.~{Montano} and S.~{Camera}.
\newblock \emph{Physics of the Dark Universe}, 46:\penalty0 101570, Dec.
  2024{\natexlab{a}}.
\newblock \doi{10.1016/j.dark.2024.101570}.

\bibitem[{Montano} and {Camera}(2024{\natexlab{b}})]{Montano2024_2}
F.~{Montano} and S.~{Camera}.
\newblock \emph{Physics of the Dark Universe}, 46:\penalty0 101634, Dec.
  2024{\natexlab{b}}.
\newblock \doi{10.1016/j.dark.2024.101634}.

\bibitem[Nasirudin et~al.(2026)Nasirudin, author2, author3, author4, and
  author5]{Nasirudin01.2026.SKA}
A.~Nasirudin et al.
\newblock In \emph{Advancing Astrophysics with the SKA -- II (AASKAII)}. 2026.
\newblock arXiv search: Report number AASKAII/Nasirudin01.

\bibitem[Navas et~al.(2024)]{ParticleDataGroup:2024cfk}
S.~Navas et~al.
\newblock \emph{Phys. Rev. D}, 110\penalty0 (3):\penalty0 030001, 2024.
\newblock \doi{10.1103/PhysRevD.110.030001}.

\bibitem[Noorikuhani and Scoccimarro(2023)]{Noorikuhani_2023}
M.~Noorikuhani and R.~Scoccimarro.
\newblock \emph{Physical Review D}, 107\penalty0 (8), Apr. 2023.
\newblock ISSN 2470-0029.
\newblock \doi{10.1103/physrevd.107.083528}.
\newblock URL \url{http://dx.doi.org/10.1103/PhysRevD.107.083528}.

\bibitem[{Norris} et~al.(2011){Norris}, {Hopkins}, {Afonso}, {Brown}, {Condon},
  et~al.]{2011PASA...28..215N}
R.~P. {Norris} et al.
\newblock \emph{\pasa}, 28\penalty0 (3):\penalty0 215--248, Aug. 2011.
\newblock \doi{10.1071/AS11021}.

\bibitem[{Novara} et~al.(2025){Novara}, {Montano}, and
  {Camera}]{Novara2025_preprint}
M.~{Novara}, F.~{Montano}, and S.~{Camera}.
\newblock \emph{arXiv e-prints}, art. arXiv:2509.08056, Sept. 2025.
\newblock \doi{10.48550/arXiv.2509.08056}.

\bibitem[Obreschkow et~al.(2009)Obreschkow, Kl\"ockner, Heywood, Levrier, and
  Rawlings]{obreschkow2009}
D.~Obreschkow et al.
\newblock \emph{\apj}, 703\penalty0 (2):\penalty0 1890--1903, Oct. 2009.
\newblock \doi{10.1088/0004-637X/703/2/1890}.

\bibitem[Padmanabhan(2003)]{Padmanabhan:2002ji}
T.~Padmanabhan.
\newblock \emph{Phys. Rept.}, 380:\penalty0 235--320, 2003.
\newblock \doi{10.1016/S0370-1573(03)00120-0}.

\bibitem[Paul et~al.(2023)Paul, Clarkson, and Maartens]{Paul:2022xfx}
P.~Paul, C.~Clarkson, and R.~Maartens.
\newblock \emph{JCAP}, 04:\penalty0 067, 2023.
\newblock \doi{10.1088/1475-7516/2023/04/067}.

\bibitem[Paul et~al.(2024)Paul, Clarkson, and Maartens]{Paul:2024uim}
P.~Paul, C.~Clarkson, and R.~Maartens.
\newblock \emph{Phys. Rev. Lett.}, 133\penalty0 (12):\penalty0 121001, 2024.
\newblock \doi{10.1103/PhysRevLett.133.121001}.

\bibitem[Peebles(1980)]{Peebles:1980yev}
P.~J. Peebles.
\newblock \emph{{The Large-Scale Structure of the Universe}}.
\newblock Princeton University Press, 11 1980.
\newblock ISBN 978-0-691-08240-0, 978-0-691-20983-8, 978-0-691-20671-4.

\bibitem[Perera et~al.(2019)]{IPTA_2019}
B.~B.~P. Perera et~al.
\newblock \emph{Mon. Not. Roy. Astron. Soc.}, 490\penalty0 (4):\penalty0
  4666--4687, 2019.
\newblock \doi{10.1093/mnras/stz2857}.

\bibitem[{Planck Collaboration: Aghanim} et~al.(2020)]{Planck:2018vyg}
{Planck Collaboration: Aghanim} et~al.
\newblock \emph{Astron. Astrophys.}, 641:\penalty0 A6, 2020.
\newblock \doi{10.1051/0004-6361/201833910}.
\newblock [Erratum: Astron.Astrophys. 652, C4 (2021)].

\bibitem[Pogosian et~al.(2010)Pogosian, Silvestri, Koyama, and
  Zhao]{Pogosian:2010tj}
L.~Pogosian, A.~Silvestri, K.~Koyama, and G.-B. Zhao.
\newblock \emph{Phys. Rev. D}, 81:\penalty0 104023, 2010.
\newblock \doi{10.1103/PhysRevD.81.104023}.

\bibitem[Pourtsidou et~al.(2013)Pourtsidou, Skordis, and
  Copeland]{Pourtsidou:2013nha}
A.~Pourtsidou, C.~Skordis, and E.~J. Copeland.
\newblock \emph{Phys. Rev. D}, 88\penalty0 (8):\penalty0 083505, 2013.
\newblock \doi{10.1103/PhysRevD.88.083505}.

\bibitem[{Rassat} and {Refregier}(2012)]{2012A&A...540A.115R}
A.~{Rassat} and A.~{Refregier}.
\newblock \emph{\aap}, 540:\penalty0 A115, Apr. 2012.
\newblock \doi{10.1051/0004-6361/201118638}.

\bibitem[{Reimberg} et~al.(2016){Reimberg}, {Bernardeau}, and
  {Pitrou}]{Reimberg_2016}
P.~{Reimberg}, F.~{Bernardeau}, and C.~{Pitrou}.
\newblock \emph{\jcap}, 2016\penalty0 (1):\penalty0 048--048, Jan. 2016.
\newblock \doi{10.1088/1475-7516/2016/01/048}.

\bibitem[Rocha and Martins(2022)]{Rocha2022}
B.~A.~R. Rocha and C.~J. A.~P. Martins.
\newblock \emph{\mnras}, 518\penalty0 (2):\penalty0 2853--2869, 11 2022.

\bibitem[{Rossiter} et~al.(2024){Rossiter}, {Camera}, {Clarkson}, and
  {Maartens}]{2024arXiv240706301R}
S.~{Rossiter}, S.~{Camera}, C.~{Clarkson}, and R.~{Maartens}.
\newblock \emph{arXiv e-prints}, art. arXiv:2407.06301, July 2024.
\newblock \doi{10.48550/arXiv.2407.06301}.

\bibitem[{Rossiter} et~al.(2026){Rossiter}, {Camera}, {Montano}, {Clarkson},
  {Karagiannis}, and {Maartens}]{2026arXiv260300244R}
S.~J. {Rossiter} et al.
\newblock \emph{arXiv e-prints}, art. arXiv:2603.00244, Feb. 2026.
\newblock \doi{10.48550/arXiv.2603.00244}.

\bibitem[{Sakr}(2026)]{2026arXiv260116899S}
Z.~{Sakr}.
\newblock \emph{arXiv e-prints}, art. arXiv:2601.16899, Jan. 2026.
\newblock \doi{10.48550/arXiv.2601.16899}.

\bibitem[{Sandage}(1962)]{sandage1962}
A.~{Sandage}.
\newblock \emph{\apj}, 136:\penalty0 319, Sept. 1962.
\newblock \doi{10.1086/147385}.

\bibitem[{Santos} et~al.(2016){Santos}, {Bull}, {Camera}, {Chen}, {Fonseca},
  et~al.]{2016mks..confE..32S}
M.~{Santos} et al.
\newblock In \emph{MeerKAT Science: On the Pathway to the SKA}, page~32, Jan.
  2016.
\newblock \doi{10.22323/1.277.0032}.

\bibitem[{Saraf} et~al.(2025){Saraf}, {Parkinson}, {Asorey}, {Hale},
  {Bahr-Kalus}, et~al.]{2025arXiv250505821S}
C.~S. {Saraf} et al.
\newblock \emph{arXiv e-prints}, art. arXiv:2505.05821, May 2025.
\newblock \doi{10.48550/arXiv.2505.05821}.

\bibitem[{Scoccimarro}(2015)]{Scoccimarro_2015}
R.~{Scoccimarro}.
\newblock \emph{\prd}, 92\penalty0 (8):\penalty0 083532, Oct. 2015.
\newblock \doi{10.1103/PhysRevD.92.083532}.

\bibitem[Seljak(2009)]{Seljak:2008xr}
U.~Seljak.
\newblock \emph{Phys. Rev. Lett.}, 102:\penalty0 021302, 2009.
\newblock \doi{10.1103/PhysRevLett.102.021302}.

\bibitem[{Shimwell} et~al.(2017){Shimwell}, {R{\"o}ttgering}, {Best},
  {Williams}, {Dijkema}, et~al.]{2017A&A...598A.104S}
T.~W. {Shimwell} et al.
\newblock \emph{\aap}, 598:\penalty0 A104, Feb. 2017.
\newblock \doi{10.1051/0004-6361/201629313}.

\bibitem[{SKA Cosmology SWG: Bacon} et~al.(2020)]{SKA:2018ckk}
{SKA Cosmology SWG: Bacon} et~al.
\newblock \emph{Publ. Astron. Soc. Austral.}, 37:\penalty0 e007, 2020.
\newblock \doi{10.1017/pasa.2019.51}.

\bibitem[Sobral-Blanco and Bonvin(2021)]{Sobral-Blanco:2021cks}
D.~Sobral-Blanco and C.~Bonvin.
\newblock \emph{Phys. Rev. D}, 104\penalty0 (6):\penalty0 063516, 2021.
\newblock \doi{10.1103/PhysRevD.104.063516}.

\bibitem[Spergel and Steinhardt(2000)]{Spergel:1999mh}
D.~N. Spergel and P.~J. Steinhardt.
\newblock \emph{Phys. Rev. Lett.}, 84:\penalty0 3760--3763, 2000.
\newblock \doi{10.1103/PhysRevLett.84.3760}.

\bibitem[{The LSST Dark Energy Science Collaboration: Mandelbaum}
  et~al.(2018)]{2018arXiv180901669T}
{The LSST Dark Energy Science Collaboration: Mandelbaum} et~al.
\newblock \emph{arXiv e-prints}, art. arXiv:1809.01669, Sept. 2018.
\newblock \doi{10.48550/arXiv.1809.01669}.

\bibitem[Tulin and Yu(2018)]{Tulin:2017ara}
S.~Tulin and H.-B. Yu.
\newblock \emph{Phys. Rept.}, 730:\penalty0 1--57, 2018.
\newblock \doi{10.1016/j.physrep.2017.11.004}.

\bibitem[Tutusaus et~al.(2023)Tutusaus, Sobral-Blanco, and
  Bonvin]{Tutusaus:2022cab}
I.~Tutusaus, D.~Sobral-Blanco, and C.~Bonvin.
\newblock \emph{Phys. Rev. D}, 107\penalty0 (8):\penalty0 083526, 2023.
\newblock \doi{10.1103/PhysRevD.107.083526}.

\bibitem[{Umeh} et~al.(2017){Umeh}, {Jolicoeur}, {Maartens}, and
  {Clarkson}]{Umeh_2017}
O.~{Umeh}, S.~{Jolicoeur}, R.~{Maartens}, and C.~{Clarkson}.
\newblock \emph{\jcap}, 2017\penalty0 (3):\penalty0 034, Mar. 2017.
\newblock \doi{10.1088/1475-7516/2017/03/034}.

\bibitem[Verbiest et~al.(2021)Verbiest, Os{\l}owski, and
  Burke-Spolaor]{Verbiest:2021}
J.~P.~W. Verbiest, S.~Os{\l}owski, and S.~Burke-Spolaor.
\newblock \emph{Pulsar Timing Array Experiments}, pages 1--42.
\newblock Springer Singapore, Singapore, 2021.
\newblock ISBN 978-981-15-4702-7.
\newblock \doi{10.1007/978-981-15-4702-7_4-1}.
\newblock URL \url{https://doi.org/10.1007/978-981-15-4702-7_4-1}.

\bibitem[{Viljoen} et~al.(2020){Viljoen}, {Fonseca}, and
  {Maartens}]{2020JCAP...09..054V}
J.-A. {Viljoen}, J.~{Fonseca}, and R.~{Maartens}.
\newblock \emph{\jcap}, 2020\penalty0 (9):\penalty0 054, Sept. 2020.
\newblock \doi{10.1088/1475-7516/2020/09/054}.

\bibitem[Weinberg(1989)]{Weinberg:1988cp}
S.~Weinberg.
\newblock \emph{Rev. Mod. Phys.}, 61:\penalty0 1--23, 1989.
\newblock \doi{10.1103/RevModPhys.61.1}.

\bibitem[{Weltman} et~al.(2020){Weltman}, {Bull}, {Camera}, {Kelley},
  {Padmanabhan}, et~al.]{2020PASA...37....2W}
A.~{Weltman} et al.
\newblock \emph{\pasa}, 37:\penalty0 e002, Jan. 2020.
\newblock \doi{10.1017/pasa.2019.42}.

\bibitem[White et~al.(2009)White, Song, and Percival]{White_2009}
M.~White, Y.-S. Song, and W.~J. Percival.
\newblock \emph{\mnras}, 397\penalty0 (3):\penalty0 1348–1354, Aug. 2009.
\newblock ISSN 1365-2966.
\newblock \doi{10.1111/j.1365-2966.2008.14379.x}.
\newblock URL \url{http://dx.doi.org/10.1111/j.1365-2966.2008.14379.x}.

\bibitem[Wolz et~al.(2026)Wolz, author2, author3, author4, and
  author5]{Wolz01.2026.SKA}
L.~Wolz et al.
\newblock In \emph{Advancing Astrophysics with the SKA -- II (AASKAII)}. 2026.
\newblock arXiv search: Report number AASKAII/Wolz01.

\bibitem[{Wu} and {Zhang}(2022)]{2022JCAP...01..060W}
P.-J. {Wu} and X.~{Zhang}.
\newblock \emph{\jcap}, 2022\penalty0 (1):\penalty0 060, Jan. 2022.
\newblock \doi{10.1088/1475-7516/2022/01/060}.

\bibitem[{Wu} et~al.(2023){Wu}, {Li}, {Zhang}, and
  {Zhang}]{2023SCPMA..6670413W}
P.-J. {Wu}, Y.~{Li}, J.-F. {Zhang}, and X.~{Zhang}.
\newblock \emph{Science China Physics, Mechanics, and Astronomy}, 66\penalty0
  (7):\penalty0 270413, July 2023.
\newblock \doi{10.1007/s11433-022-2104-7}.

\bibitem[Yahya et~al.(2015)Yahya, Bull, Santos, Silva, Maartens, Okouma, and
  Bassett]{yahya2015}
S.~Yahya et al.
\newblock \emph{\mnras}, 450\penalty0 (3):\penalty0 2251--2260, 05 2015.

\bibitem[{Yamamoto} et~al.(2006){Yamamoto}, {Nakamichi}, {Kamino}, {Bassett},
  and {Nishioka}]{Yamamoto_2005}
K.~{Yamamoto} et al.
\newblock \emph{\pasj}, 58:\penalty0 93--102, Feb. 2006.
\newblock \doi{10.1093/pasj/58.1.93}.

\bibitem[{Yoo} and {Desjacques}(2013)]{2013PhRvD..88b3502Y}
J.~{Yoo} and V.~{Desjacques}.
\newblock \emph{\prd}, 88\penalty0 (2):\penalty0 023502, July 2013.
\newblock \doi{10.1103/PhysRevD.88.023502}.

\bibitem[Yoo et~al.(2009)Yoo, Fitzpatrick, and Zaldarriaga]{Yoo:2009au}
J.~Yoo, A.~L. Fitzpatrick, and M.~Zaldarriaga.
\newblock \emph{Phys. Rev. D}, 80:\penalty0 083514, 2009.
\newblock \doi{10.1103/PhysRevD.80.083514}.

\bibitem[{Zaroubi} and {Hoffman}(1993)]{Zaroubi_1993}
S.~{Zaroubi} and Y.~{Hoffman}.
\newblock \emph{arXiv e-prints}, art. astro-ph/9311013, Nov. 1993.
\newblock \doi{10.48550/arXiv.astro-ph/9311013}.

\bibitem[Zhang et~al.(2007)Zhang, Liguori, Bean, and Dodelson]{Zhang:2007nk}
P.~Zhang, M.~Liguori, R.~Bean, and S.~Dodelson.
\newblock \emph{Phys. Rev. Lett.}, 99:\penalty0 141302, 2007.
\newblock \doi{10.1103/PhysRevLett.99.141302}.

\end{thebibliography}
